\documentclass[11pt,twocolumn]{aastex63}
\usepackage{url}
\usepackage{gensymb}
\usepackage{textcomp}
\usepackage{xspace}
\usepackage{natbib,amsmath} \usepackage{graphicx}
\usepackage{hyperref} \usepackage{float} \usepackage{subfigure}
\usepackage{tabularx}
\providecommand{\e}[1]{\ensuremath{\times 10^{#1}}}
\usepackage{bm}
\newcommand{\um}{\textmu m\xspace}

\newcommand{\ang}{\AA\xspace}

\newcommand*\chem[1]{\ensuremath{\mathrm{#1}}}
\newcommand{\mean}[1]{\langle #1\rangle}

\newcommand{\cm}{\mathrm{cm}}
\newcommand{\g}{\mathrm{g}}
\newcommand{\K}{\mathrm{K}}  
\newcommand{\eV}{\mathrm{eV}}

\newcommand{\s}{\mathrm{s}}
\newcommand{\yr}{\mathrm{yr}}
\newcommand{\p}{\mathrm{p}}
\newcommand{\mathang}{\ensuremath{\mathrm{\ang}}}
\newcommand{\cloudy}{\texttt{CLOUDY}\xspace}
\usepackage{array}

\begin{document}
\title{No escaping helium from 55 Cnc e \footnote{ The data presented herein were obtained at the W. M. Keck Observatory, which is operated as a scientific partnership among the California Institute of Technology, the University of California and the National Aeronautics and Space Administration. The Observatory was made possible by the generous financial support of the W. M. Keck Foundation.}}

\correspondingauthor{Michael Zhang}
\email{mzzhang2014@gmail.com}

\author[0000-0002-0659-1783]{Michael Zhang}
\affiliation{Department of Astronomy, California Institute of Technology, Pasadena, CA 91125, USA}

\author[0000-0002-5375-4725]{Heather A. Knutson}
\affiliation{Division of Geological and Planetary Sciences, California Institute of Technology}

\author[0000-0002-6540-7042]{Lile Wang}
\affiliation{Flatiron Institute}

\author[0000-0002-8958-0683]{Fei Dai}
\affiliation{Division of Geological and Planetary Sciences, California Institute of Technology}

\author[0000-0002-9584-6476]{Antonija Oklopcic}
\affiliation{Anton Pannekoek Institute for Astronomy, University of Amsterdam, Science Park 904, 1098 XH Amsterdam, The Netherlands}

\author[0000-0003-2215-8485]{Renyu Hu}
\affiliation{Jet Propulsion Laboratory}
  
\begin{abstract}
We search for escaping helium from the hot super Earth 55 Cnc e by taking high-resolution spectra of the 1083 nm line during two transits using Keck/NIRSPEC.  We detect no helium absorption down to a 90\% upper limit of 250 ppm in excess absorption or 0.27 m\ang  in equivalent width.  This corresponds to a mass loss rate of less than $\sim10^9$ g/s assuming a Parker wind model with a plausible exosphere temperature of 5000-6000 K, although the precise constraint is heavily dependent on model assumptions.  We consider both hydrogen- and helium-dominated atmospheric compositions, and find similar bounds on the mass loss rate in both scenarios.  Our hydrodynamical models indicate that if a lightweight atmosphere exists on 55 Cnc e, our observations would have easily detected it.  Together with the non-detection of Lyman $\alpha$ absorption by \cite{ehrenreich_2012}, our helium non-detection indicates that 55 Cnc e either never accreted a primordial atmosphere in the first place, or lost its primordial atmosphere shortly after the dissipation of the gas disk.
\end{abstract}

% #####################################################################

\section{Introduction}
\label{sec:introduction}
The observed radius distribution of sub-Neptune-sized planets is bimodal, with peaks at $<$ 1.5 R$_\earth$ and 2-3 R$_\earth$ \citep{fulton_2017}.  This bimodality can be explained if the observed population of sub-Neptune-sized planets formed with several M$_\earth$ rocky cores and hydrogen-rich atmospheres, which were then stripped away from the most highly irradiated planets (i.e. \citealt{lopez_2013,ginzburg_2018}; see \citealt{owen_2019} for a literature review).  The high inferred core densities in these models argue strongly for formation inside the ice line, and the semi-major axis distribution of this population of planets is well-matched by in situ formation models \citep{lee_2017}.  While the prevailing evidence at the moment appears to favor in situ (or at least nearby) formation, the arguments proposed to date are by no means definitive as they rely on indirect model-based inferences.  If the mass loss models used to simulated the observed radius distribution are incomplete or rely on incorrect assumptions, our conclusions about these planets may be incorrect as well. It is therefore important to have observational data to nail down theoretical models.

In this study, we focus on one of the most observationally favorable transiting super-Earths currently known: 55 Cnc e.  55 Cnc is a binary system with a K0 main sequence star and a M dwarf companion, separated by 1000 AU.  The primary is a bright (V=5.95) star whose activity and rotation rate indicate it is very old, probably around 10 Gyr.  It has at least 5 planets of various sizes and orbital distances, including e, a superheated $R=1.88 \pm 0.03 R_\earth$ \citep{bourrier_2018} super-Earth with a period of 0.74 days and the only planet known to transit.  The mean density of e as derived from radial velocities and transit depths is 6.7 g/cm\textsuperscript{3}, suggesting a rocky interior with an atmosphere contributing up to a few percent of the planet radius \citep{bourrier_2018}, although it is also consistent with a small iron core with a silicate mantle and no water or gas layer.  At first glance 55 Cnc e seems unlikely to host a primordial hydrogen and helium rich atmosphere as it has a relatively high equilibrium temperature and correspondingly high predicted escape rate \citep{valencia_2010}.  However, there are large theoretical uncertainties in mass loss models, driven in part by the uncertain X-ray and extreme ultraviolet spectrum of the star \citep{owen_2019}.  It has also been suggested that this planet might have a lava ocean on its dayside that could outgas enough trapped hydrogen to form a thick secondary atmosphere \citep{chachan_2018}.  

Current observational constraints on 55 Cnc e's atmosphere are also conflicting.  Some observations, such as the non-detection of hydrogen Lyman $\alpha$ absorption by \cite{ehrenreich_2012} or the variations in infrared emission reported by \cite{demory_2016a}, seem to indicate that it is unlikely to host a substantial hydrogen-rich atmosphere.  Other observations suggest the opposite, including the planet's small measured day-night temperature gradient \citep{demory_2016b}, the tentative evidence for sodium and calcium absorption \citep{ridden-harper_2016}, and the detection of a strong absorption feature in the planet's 1.1--1.7 $\mu m$ transmission spectrum from \emph{Hubble Space Telescope} (\emph{HST}) WFC3, which has been attributed to HCN \citep{tsiaras_2016}.  Regardless of its source, the large amplitude of the WFC3 absorption feature can only be matched by a hydrogen- or helium-dominated atmosphere, because an atmosphere dominated by heavier elements would have a much smaller scale height and correspondingly weaker absorption features during transit. If 55 Cnc e does have a significant low mean molecular weight atmosphere, its ability to retain this atmosphere in the face of ongoing mass loss would place strict constraints on the magnitude of relevant mass loss processes, with correspondingly wide-reaching implications for our understanding of the overall population of short-period super-Earths and sub-Neptunes.  

In this paper, we use the helium 1083 nm metastable triplet \citep{oklopcic_2018} to search for evidence of helium outflow from 55 Cnc e.  Unlike previous Lyman $\alpha$ observations, this helium triplet is easily accessible using ground-based telescopes, and has been used to detect the extended atmospheres of multiple exoplanets.  Most detections have been around Jupiter radius planets such as WASP-107b (i.e. \citealt{spake_2018}), with the 4 R$_\earth$ GJ 3470b  being the smallest planet with a detection to date \citep{ninan_2020,palle_2020}.  For 55 Cnc e, our observations of the 1083 nm helium line supplements the \cite{ehrenreich_2012} Lyman alpha observations in shaping our understanding of the exosphere.  The \cite{ehrenreich_2012} observations took place at a stellar activity minimum, when the star's X-ray flux was 2--3 times lower than during our observations (see Section \ref{sec:discussion}); this decreased X-ray flux might have suppressed the mass loss rate during these observations.  In addition, some studies have proposed that mass loss over many Gyr can preferentially remove hydrogen from the atmosphere of a small planet while leaving helium behind \citep{hu_2015,malsky_2020}.  A helium dominated atmosphere would be consistent with the molecular weight inferred from the HST transit spectrum \citep{tsiaras_2016}.  Such an atmosphere might be undetectable in Lyman $\alpha$, but easy to detect using metastable helium.  Finally, because Lyman $\alpha$ is such a strong line, the observed signal is dependent on the behavior of the diffuse exosphere far from the planet.  The atmospheric absorption signal in the helium 1083 nm triplet is typically much weaker than the absorption signal in the Lyman-$\alpha$ line, but IR measurements are significantly more precise than UV measurements due to the much higher photon flux and (unlike Lyman $\alpha$) we are able to observe the cores of the lines.  This means that helium observations are sensitive to gas at smaller radii and with lower outflow velocities than Lyman $\alpha$ observations, making it easier to compare to mass loss models.  
It is for these three reasons--the differing stellar XUV irradiation, the possibility of a helium-dominated atmosphere, and the complementary sensitivities--that observations of the helium line are meaningful even for small planets with existing Lyman alpha non-detections.

We describe our observations in Section 2, data reduction pipeline in Section 3, analysis in Section 4, outflow models in Section 5, and scientific implications in Section 6.

\section{Observations}
We observed two transits of 55 Cnc e in Y band using the upgraded NIRSPEC instrument on Keck \citep{martin_2018}: one on December 4, 2019 at 13:56 UTC (barycentric), and one on December 18, 2019 at 13:48 UTC (barycentric).  These observations used the 0.288 x 12 arcsec slit, giving NIRSPEC a resolution of 37,500, with a FWHM sampling of 3 pixels.  All observations were performed with 60 second exposure times in an ABBA nod pattern to facilitate background subtraction.  Details of the observations are given in Table \ref{table:nirspec_data}.

\begin{table}[ht]
  \centering
  \caption{Keck/NIRSPEC observations}
  \begin{tabular}{c c c}
  \hline
  	  Parameter & 12/04/19 (UTC) & 12/18/19 (UTC)\\
      \hline
      Obs. duration & 3.2 h & 3.7 h\\
      Number of obs & 107 & 136\\
      Obs before transit & 23 & 40\\
      Obs during transit & 51 & 56\\
      Obs after transit & 33 & 40\\
      Airmass before transit & 1.1 & 1.04\\
      Airmass during transit & 1.02 & 1.01\\
      Airmass after transit & 1.03 & 1.11\\
      Obs efficiency & 56\% & 61\%\\
      Avg. SNR$^*$ & 660 & 640\\
      \hline
  \end{tabular}
  \label{table:nirspec_data}
  \tablecomments{$^*$Average SNR is calculated per spectral pixel per exposure, based on the estimated error from optimal extraction.}
\end{table}

Observation conditions were stable with good seeing during the first night.  The seeing was worse and more variable during the second night, decreasing the per-pixel count rate on the detector and making it possible to use a more efficient observing strategy of 20 second subexposures with 3 co-adds instead of 15 second subexposures with 4 co-adds.  The marginally increased observing efficiency could not fully compensate for the higher seeing, leading the second night to have a marginally worse SNR per spectrum.

For unknown reasons and at unpredictable times, the telescope would fail to nod. We encountered this issue on each of our four half-nights with NIRSPEC--two in April 2019, and two in December 2019. Sometimes it would not move at all; at other times, it would jump out of the slit, either by nodding too far, or by adding a perpendicular component to the nod.  When this happened, we would wait for the exposure to finish (NIRSPEC does not allow us to stop during the middle of an exposure), re-center the star, and re-start the nod sequence.  We then discard the exposure, together with the previous exposure if it is necessary for background subtraction.  On 12/04, nodding failed once during the transit and we lost 4 exposures (because we did not notice the failure immediately), corresponding to 7 minutes of observation time.  On 12/18, nodding failed twice during the transit and once after the transit, resulting in a loss of 6 exposures.  Other than these nodding failures, our observing sequence is continuous on both nights.

\section{Data reduction}
We calibrated the raw images and extracted 1D spectra for each order using a custom Python pipeline designed for the upgraded NIRSPEC.  We describe each step of this process below. 

\subsection{Crosstalk removal}
In its current configuration NIRSPEC is divided into 32 readout channels, with 64 rows per channel.  The rows of adjacent channels are read simultaneously, but in reverse order: row 0 of channel 0 should be read at the same time as row 63 of channel 1.  The simultaneous readout causes crosstalk signals between each pair of channels.  We follow the approach described in \cite{george_2020} and initially assume that the crosstalk signal $X_{k,j,i}$ in NIRSPEC is proportional to the derivative of the signal in the source channel: 

\begin{align}
X_{k,j,i} = a_{k,j}(S_{k,i} - S_{k,i-1}),
\label{eq:crosstalk}
\end{align}

where channel $k$ is the source of the crosstalk, channel $j$ is the destination, $i$ is the row number, and $a$ is a 32$\times$32 matrix of scaling factors whose elements are of the order 400 ppm.  The crosstalk introduced is therefore on the order a few hundred electrons for signals on the order of 1,000,000 electrons, but varies widely for different $k,j$ pairs.

Unfortunately, there are no calibration data for NIRSPEC that can help to characterize the crosstalk, as we require a high flux derivative along the spatial axis.  We therefore use archival NIRSPEC observations of 55 Cnc in L band from April 2 and 16, 2019 to measure the crosstalk matrix $a$.  On each night, we took $\sim$50 minute-long exposures, each with very high flux ($\sim$300,000 electrons per spectral pixel).  To measure the crosstalk matrix $a$, we masked out pixels with a count greater than 70 ADU (200 electrons) and used a linear least squares fit to estimate $a$ and the uncertainty on $a$.  The uncertainty is high where the source channel is un-illuminated and/or the target channel has few un-illuminated pixels, and low where the source channel is illuminated but the target channel is not.  We perform a weighted average over the estimates of $a$ from each exposure on each night to obtain our best estimate for the matrix.

\begin{figure}[ht]
  \centering \subfigure {\includegraphics
    [width=0.5\textwidth]{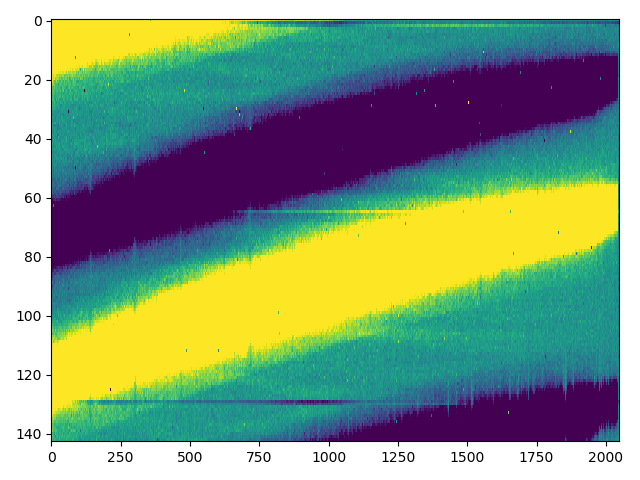}}\qquad
  \subfigure {\includegraphics
    [width=0.5\textwidth]{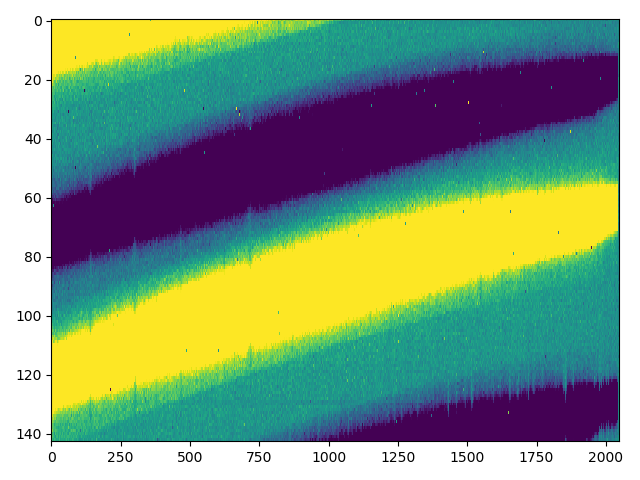}}
    \caption{A portion of a raw A-B frame, showing the order containing the helium line.  Top: no calibration corrections.  Bottom: crosstalk subtracted.  Notice that the subtle horizontal ripple pattern is gone, as are the two sharp horizontal lines.}
\label{fig:crosstalk}
\end{figure}

Relying on science data to measure the crosstalk matrix is less than ideal because channels that have no trace do not produce a measurable crosstalk signal in other channels; the trace dominates the crosstalk from other channels in channels where the trace is nearly horizontal and close to the middle.  Nevertheless, a visual inspection indicates that this imperfect approach to crosstalk subtraction is sufficient to remove the crosstalk signal from our data (Figure \ref{fig:crosstalk}).

We find that for NIRSPEC, Equation \ref{eq:crosstalk} is less accurate when j and i are of different parity.  By carefully examining the crosstalk patterns, we found that the crosstalk signal is proportional not to $S_{k,i} - S_{k,i-1}$, but to $S_{k,i+2} - S_{k,i+1}$.  In physical terms, this means that the 2nd rows of odd channels (not the 0th rows) are read at the same time as the 63rd rows of even channels.  Similarly, the 1st row corresponds to the 62nd, the 2nd to the 61st, and so on, until the 63rd row of odd channels are read simultaneously with the 2nd rows of even channels.  Therefore, rows 0 and 1 of every channel are being read out when their sister channels of opposite parity are not being read out.

We believe this mismatch in the readout pattern causes another detector artifact: the anomalous rows, seen in Figure \ref{fig:crosstalk} as horizontal lines.  Anomalous rows occur when the row number modulo 128 is equal to 0, 1, or 64.  The first two correspond to rows 0 and 1 of even channels; the last corresponds to row 0 of odd channels.  We speculate that the anomalous rows could be caused by crosstalk: when row 0 or 1 is being read out, the corresponding readout line in opposite parity channels could be carrying signals of much higher amplitude than image data, causing much higher crosstalk than normal.

In the absence of photons, the anomalous rows are mostly, but not entirely, consistent across the entire image.  Thus, all 32 rows divisible by 128 are similar; all rows whose remainder is 1 when divided by 128 are similar; and all rows whose remainder is 64 when divided by 128 are similar.  We create a template of these 3 different categories of anomalous rows by identifying regions of the detector which received few photons, either because they are in between orders, or because they are within an order but far from the trace.  At each column, the template is equal to the median of the rows fitting this criterion.  The template is subtracted from all relevant rows.  This method works very well, but not perfectly--a faint hint of under-subtraction can be seen in Figure \ref{fig:crosstalk} (bottom) for row 128.

\subsection{Image calibration}
\label{subsec:calibration}
On the first (second) night, we took 80 (63) flat fields, each with an exposure time of 4.4 s per co-add and 20 co-adds.  In total, we collected 750 million (590 million) electrons per pixel, ensuring that the photon error in the flat fields is far below the photon noise in the observational data.  To calibrate the flats, we took 19 (3) darks, each with an exposure time of 4.4 s per co-add and 20 co-adds.  We create a master dark by median stacking the individual dark frames.  In each individual dark, pixels that deviate from the image-wide mean by more than $5\sigma$ are marked as bad pixels.  Pixels marked as bad in more than half of the individual darks are marked as bad in the master dark.

We create a master flat by taking the median of the individual flats.  Prior to combining, we subtract the crosstalk from each flat frame using the algorithm described in the previous subsection, subtract the master dark, and divide by the median flux.  We identify the order containing the helium line and mask out everything else. as there is no need to extract spectra from other orders.  The relevant order is fitted with a polynomial that is 5th order with in x and 3rd order in y, then divided by the polynomial, in order to bring the values of all good pixels close to 1.  We then create a mask of bad pixels, including both the bad pixels identified in the master dark, and the pixels with a flat value lower than 0.5 or higher than 1.5.  The final master flat, together with the bad pixel mask, are saved in a FITS file.

Finally, we correct the raw science frames using the master flat.  For each $A_1B_1B_2A_2$ nod, we create four difference images: $A_1 - B_1$, $B_1 - A_1$, $B_2 - A_2$, and $A_2 - B_2$.  Each difference image is crosstalk subtracted, multiplied by the gain of g=2.85 e/ADU, and divided by the master flat F.  We next construct a variance image for each differenced frame, which indicates the uncertainty in the measured flux at each individual pixel location. This variance image includes the photon noise from the star, the sky background, and the detector read noise.  The background $b$, read noise $N_R$, and total variance $V$ are computed as follows:

\begin{align}
    b = g\frac{A+B-|A-B|}{F}     \label{eq:bkd}\\
    N_R = \sqrt{2}\sqrt{\frac{N_{\rm coadds}}{N_{\rm reads}}} N_{R,0}\\
    V = g\frac{|A| + |B|}{F^2} + N_R^2,
\end{align}

where $N_{R,0} = 44 e^-$, $N_{\rm reads}=4$ for all of our observations, $g=2.85$ $e^-$/ADU is the gain, and $N_{\rm coadds}=4$ on our first night and $N_{\rm coadds}=3$ on most of our second night.  The iterative bad pixel algorithm in REDSPEC 3.0\footnote{\url{https://www2.keck.hawaii.edu/inst/nirspec/redspec.html}} \citep{kim_2015}, which we ported to Python, is used to identify and repair bad pixels.  This algorithm identifies bad pixels with a variant of local sigma clipping.  In total, it typically identifies and repairs 1000 hot pixels and 1000 cold pixels in a given image.  Since the identification is limited to the $~\sim$100 pixel tall order containing the helium line, 2000 bad pixels represents 1\% of all pixels.  The bad pixels identified by the algorithm are combined with the bad pixels identified in the master dark and master flat to create a master bad pixel mask.  The difference image, variance image, background image, and bad pixel mask are all saved in a FITS file.

\subsection{Optimal extraction}
\label{subsec:opt_extract}
After calibrating the images, we extract the 1D spectrum from each 2D spectral trace.  We first determine the position of the trace in each image.  For every column (corresponding to one wavelength), we fit a Gaussian to the pixel values to estimate the trace position  We then fit a 5th order polynomial to the trace locations as a function of column number.  The residuals in this final fit are typically smaller than 0.01 pixels.  Accuracy is not paramount because we only use the trace to identify regions of the image very far from the trace, in order to mask them out.

After determining the position of the trace, we perform optimal extraction using a variant of the method described in \cite{horne_1986}.  The original \cite{horne_1986} algorithm assumes that the wavelength axis is aligned with the columns, and the spatial axis is aligned with the rows.  Unfortunately, on NIRSPEC neither the wavelength nor spatial axes are aligned with either the rows or the columns.  The axes are also not perpendicular to each other, and neither axis is straight, as can be seen in Figure \ref{fig:crosstalk}.  The typical way around this problem is to rectify the order spatially and spectrally by interpolation onto a rectilinear grid.  This works well enough for low SNR data, but for our exceptionally high SNR data it introduces aliasing artifacts at the 0.1--1\% level.

We instead use a rolling window approach where we assume that each column corresponds to one wavelength--a reasonable assumption, given that most of the flux is concentrated within 5 pixels of the center of the trace.  Regions more than 15 pixels from the center of the trace are masked out to avoid interference from neighboring orders or cosmic rays.  For each column, we take a window 81 pixels wide and 72 pixels tall, and fit the profile for every row.  We chose 81 pixels as the width because we want the window to be small enough for the trace to deviate vertically by less than one pixel, but big enough to fit the profile accurately. We chose a 72 pixel window in the $y$ (cross-dispersion) direction because the trace is 50 pixels lower on the left end of the detector than on the right, and we needed a window large enough to encompass not just the center of the trace, but also the wings across the full width of the image.

In \cite{horne_1986}, the spatial profile is estimated by dividing each column by its sum.  This runs into problems in regions of high telluric absorption, where the spectral flux approaches zero.  We therefore fit the product of the spectrum and a Chebyshev polynomial model of the profile directly to the observed data.  Mathematically, if $i$ represents the order of the Chebyshev polynomial $T_i$ and $j$ is the column number, we look for the $x$ that minimizes $Ax=b$ where $A$ and $b$ are:

\begin{align}
    A_{i,j} &= S_j T_i(\frac{j-N/2}{N}) / \sigma_j\\
    b_j &= I_j / \sigma_j,
\end{align}
$N=72$ is the size of the window and $\frac{j-N/2}{N}$ normalizes the column numbers to range from -1 to +1.  After obtaining the least squares solution to $x$, the profile can be computed as:

\begin{align}
    P_j = \sum_{i=0}^5 x_i T_i(\frac{j-N/2}{N}),
\end{align}
where 5 is the the maximum order of Chebyshev polynomials we fit.

We note that this is mathematically equivalent to fitting monomials.  If we had replaced $T_i(y)$ with $y^i$ in all the equations above, we would arrive at an identical $P_j$.  The advantage of Chebshev polynomials comes from the numerical stability and robustness of the linear algebra solver.  $y^i$ becomes extremely small when $i$ is large for all but the boundary values -1 and 1, causing numerical problems, whereas the Chebshev polynomials are well behaved for every order.  The Chebshev polynomials are also mutually orthogonal; the monomials are not.  In practice, it appears that neither of these matter for our low order profile fit, but since Chebysheb fitting is theoretically superior, we adopt it as our preferred form of polynomial fitting throughout our code.

\begin{figure*}[ht]
  \includegraphics
    [width=\textwidth]{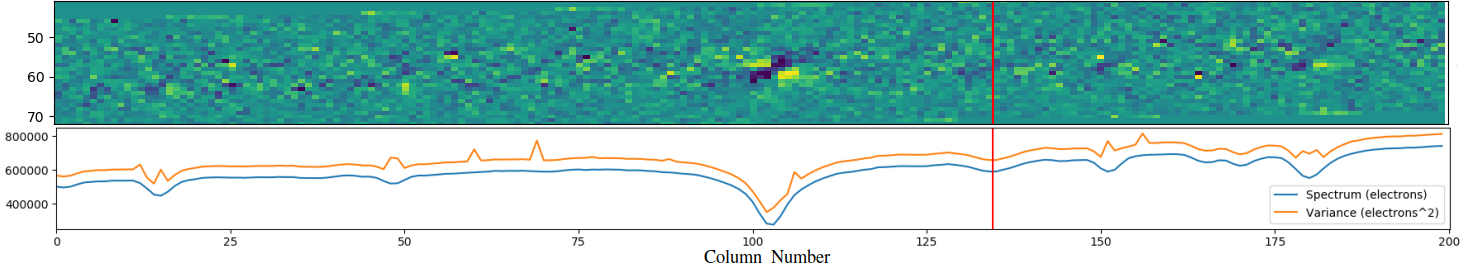}
    \caption{Top: the residuals of optimal extraction, scaled to the standard deviation of each pixel and clipped from -5 to 5$\sigma$.  If optimal extraction worked perfectly, this image would consist of Gaussian-distributed noise with a mean of 0 and a standard deviation of 1.  The four-leaf clover pattern in the middle of the residuals image corresponds to a deep stellar line at 1.0830 um.  The position of the helium 1.08333 um line is marked in red.  Although a stellar helium line is present it is much shallower than the 1.0830 um line, and there does not appear to be any strong residual structure in this region.  Bottom: optimally extracted spectrum with accompanying variance.  The spikes in variance are due to masks on bad pixels. }
\label{fig:clovering}
\end{figure*}

We make several additional modifications to the \cite{horne_1986} algorithm.  The background cannot be estimated from the difference image alone, so we instead use the estimate from Equation \ref{eq:bkd}.  Instead of identifying bad pixels solely from the image using sigma clipping, we start with the bad pixel mask from the calibration stage (Subsection \ref{subsec:calibration}) and identify additional bad pixels using the image.  Finally, we use an extremely high threshold $\sigma_{clip}=12$ to reject bad pixels during the optimal extraction iterations.  We require this high threshold because of a phenomenon that we observe with our high signal-to-noise data: the four-leaf clover pattern in the residuals image, shown in Figure \ref{fig:clovering}.  This pattern occurs wherever the spectrum is changing rapidly--namely, in the vicinity of deep stellar or telluric lines.  The pattern traces the cross derivative $\frac{dF}{dxdy}$, and is more pronounced where the trace is highly tilted with respect to the horizontal.  This clover pattern indicates that optimal extraction does not give optimal results in the vicinity of deep lines.  Regions without deep lines--including the vicinity of the 1.08333 \um helium line--are minimally affected.

We initially thought that the clover pattern was due to the misalignment between the image axes and the wavelength/spatial axes, or due to the non-orthogonality between the wavelength and spatial axes.  We tried many different optimal estimation schemes that do not make the rectilinear assumptions of \cite{horne_1986}, including \cite{marsh_1989} and an unpublished algorithm\footnote{\url{https://code.obs.carnegiescience.edu/Algorithms/ghlb/at_download/file}}, in an attempt to get rid of the clovers.  However, we eventually realized that the clover pattern is caused by the inseparability of the 2D PSF.  The optimal extraction algorithm assumes that $PSF_{\lambda,x} = S(\lambda)P(x)$ for some spectrum $S(\lambda)$ and profile $P(x)$.  A 2D Gaussian with its axes aligned with the wavelength and spatial axes is separable, but a tilted 2D Gaussian is not.  \cite{bolton_2010} simulate spectra under the assumption of PSF non-separability, apply optimal extraction to the simulated spectra, and obtain clover patterns strikingly similar to ours (see their Figure 1, right).  The non-separability of the PSF means that only algorithms like spectro-perfectionism \citep{bolton_2010} that adopt a 2D PSF can avoid the clovers.  Spectro-perfectionism is not necessary for our work here because the helium line is not in a region where the spectrum varies rapidly.

\subsection{Wavelength solution}
After extracting the 1D spectra, we determine the wavelength solution for each spectrum.  We first create a template containing stellar and telluric lines at known wavelengths, with the stellar lines shifted to account for Earth's velocity relative to the star on that night.  We adopted a $T_{eff}=5200 K$, $log(g)=4.5$, and [M/H]=0.0 PHOENIX model spectrum \citep{husser_2013} for the star.  We used a telluric transmission spectrum from the Gemini website\footnote{https://www.gemini.edu/observing/telescopes-and-sites/sites}, which assumes a precipitable water vapor of 1.6 mm and an airmass of 1.5.  This spectrum was calculated using ATRAN \citep{lord_1992}.  We multiply the stellar spectrum by the telluric transmission and downsample to instrumental resolution (R=37,500) to get our final template.

We parameterize the wavelength solution as a third-order polynomial function of the normalized column number, $j'=\frac{j-N/2}{N}$:

\begin{align}
    \lambda(j') = \sum_{i=0}^3 c_i T_i(j'),
\end{align}
where $T_i$ are the Chebyshev polynomials.  Similarly, we parameterize the continuum as a fifth-order polynomial of $j'$, and multiply it by the template.  We then use \texttt{scipy}'s differential evolution minimizer to minimize $\chi^2$, which we calculate as the difference between the observations and the continuum-adjusted template interpolated using the proposed wavelength solution coefficients.  A visual inspection of the resulting fit indicates that the line positions match to better than a pixel.

\section{Analysis}
After extracting the 1D spectra, we place the data from each night on a uniform wavelength grid and remove signals not related to the planet.  This includes instrumental effects like detector fringing, telluric absorption lines, and absorption lines from the star itself.  We then shift each spectrum into the planetary rest frame and quantify the amount of excess absorption in the helium triplet during transit.  We describe each step of this process in detail below.

\subsection{Better ephemeris}
Due to the high radial acceleration of the planet during transit, an accurate ephemeris is necessary to shift spectra into the planet frame.  We calculate an updated ephemeris by combining the epoch derived by \cite{demory_2016a} using 4 Spitzer 4.5 \um transits, the epoch derived by \cite{sulis_2019} using 143 MOST transits, and 35 individual transit timings from TESS.  We obtain the TESS transit timings using a procedure similar to that described in \citet{Dai}. In short, we downloaded the {\it TESS} photometry from the Mikulski Archive for Space Telescopes (MAST). Using the archival ephemeris, we isolated data within a wide window of three times the archival transit duration. We fitted each individual transit with the \texttt{BATMAN} \citep{Kreidberg} package and a quadratic function of time to account for local stellar variability.  We adopted quadratic limb darkening and imposed Gaussian priors
with widths of 0.3 centered around the theoretical
values from {\tt EXOFAST}\footnote{\url{astroutils.astronomy.ohio-state.edu/exofast/limbdark.shtml}.} \citep{Eastman2013}. We then fitted all {\it TESS} transits globally after removing the stellar variability component. With this global model as a template, we revisit each individual transits allowing only the mid-transit time and local quadratic function to vary. This process is iterated a few times until convergence. Finally, the individual {\it TESS} transit epochs are fitted together with the archival transit times. The updated ephemeris is $P=0.73654604 \pm 1.6 \times 10^{-7}$d and $T_0=2458723.38328 \pm 0.00014$ (BJD\textsubscript{TDB}) with negligible covariance C=-$10^{-14}$d.  The covariance C is defined such that the prediction error at epoch E from this ephemeris is:

\begin{align}
    \sigma_{T}^2 = \sigma_{T_0}^2 + E^2\sigma_P^2 + 2CE,
\end{align}

We chose the initial epoch $T_0$ to make C as close to 0 as possible, allowing the last term to be ignored.  The prediction error evaluates to 13 seconds at the time of our December 2019 observations.

\subsection{Making a spectral grid}
The first step is to linearly interpolate all spectra on a given night onto a common wavelength grid.  We choose a wavelength grid spanning 1.0826 to 1.0840 \um, which encompasses the locations of the three lines in the helium triplet. We select a resolution of 110,000 for our grid, approximately matching the native pixel resolution.  The process of interpolation introduces covariances between adjacent wavelength bins, artificially smoothing out the interpolated spectrum.  We keep track of the covariance matrix during our analysis, and eventually use it to calculate likelihoods.

\begin{figure}[ht]
  \centering \subfigure {\includegraphics
    [width=0.5\textwidth]{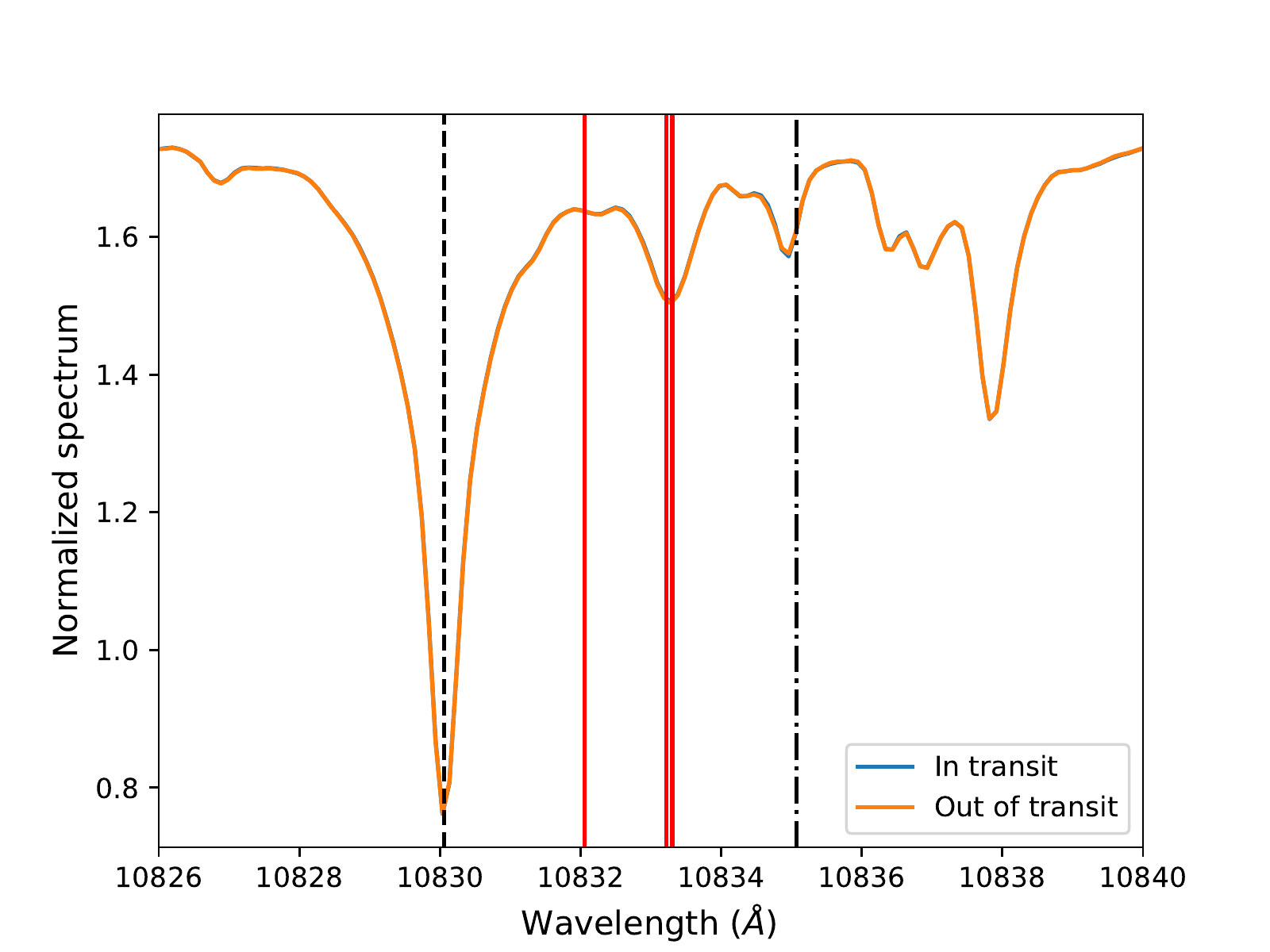}}\qquad
  \subfigure {\includegraphics
    [width=0.5\textwidth]{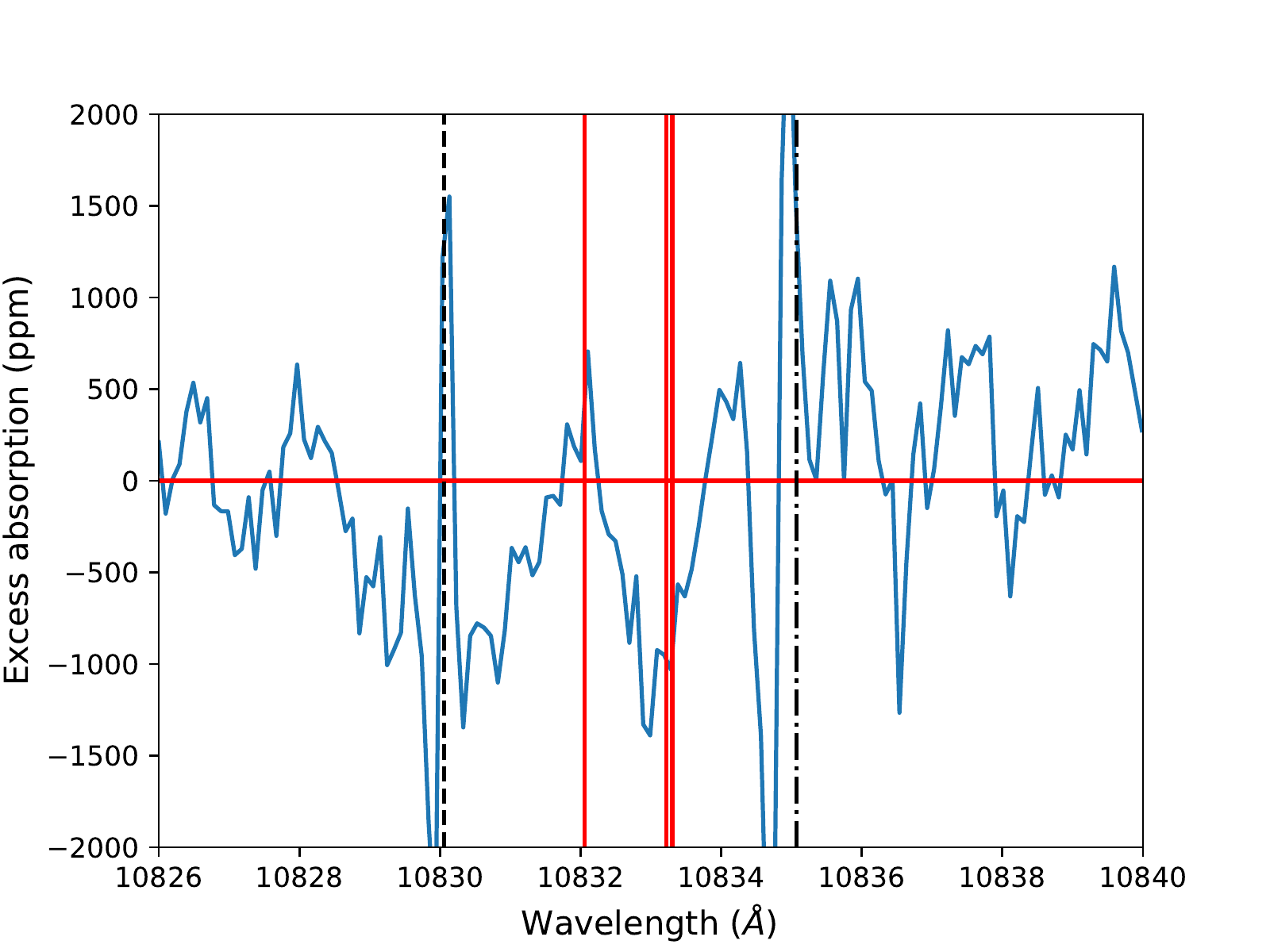}}
    \caption{Top: in-transit vs. out-of-transit spectra in the stellar rest frame, calculated using data from both nights.  Bottom: excess absorption ($F_{in}/F{out} - 1$), plotted in the stellar frame; fringing has been removed with a bandstop filter, but additional corrections for telluric and instrumental effects described in Subsection \ref{subsec:residuals} have not been applied. The red vertical lines mark the locations of the three helium lines.  The dashed vertical black lines represent strong stellar lines, while the dash-dotted vertical line is a telluric water line.  The most prominent features in the excess absorption plot are the continuum variation, and the sudden spikes/dips at the position of the strong lines.  The latter is due to the poor performance of optimal extraction in the vicinity of strong lines.  Helium absorption would manifest as a spike in the vicinity of the red vertical lines, which is not seen.}
\label{fig:stellar_frame_abs}
\end{figure}

We do an initial check for helium absorption during the transit by dividing all spectra from both nights into two categories, in-transit and out-of-transit.  We take the mean of the in-transit spectra to create a master in-transit spectrum, and the mean of the out-of-transit spectra to create a master out-of-transit spectrum.  We then calculate the excess absorption in the stellar frame as $F_{in}/F_{out} - 1$.  The two master spectra and the excess absorption are plotted in Figure \ref{fig:stellar_frame_abs}.  There does not appear to be any detectable increase in absorption during the transit at the position of the helium line.

\subsection{Residuals image}
\label{subsec:residuals}
Since we are interested in fractional changes in the spectrum, we take the natural log of the spectral grid and subtract the mean of every row and column, producing what we call the residuals image.  Every pixel in the residuals image approximately represents the fractional flux change at that epoch and wavelength from the mean spectrum.  Taking the natural log also has the advantage of linearizing the effect of changing airmass.  The observed flux is roughly $F(\lambda) = F_{vac}(\lambda)e^{-\alpha z}$ where z is the airmass, so $\ln{F(\lambda)} = \ln{F_{vac}(\lambda)} - \alpha z$. 

\begin{figure}[ht]
  \includegraphics
    [width=0.45\textwidth]{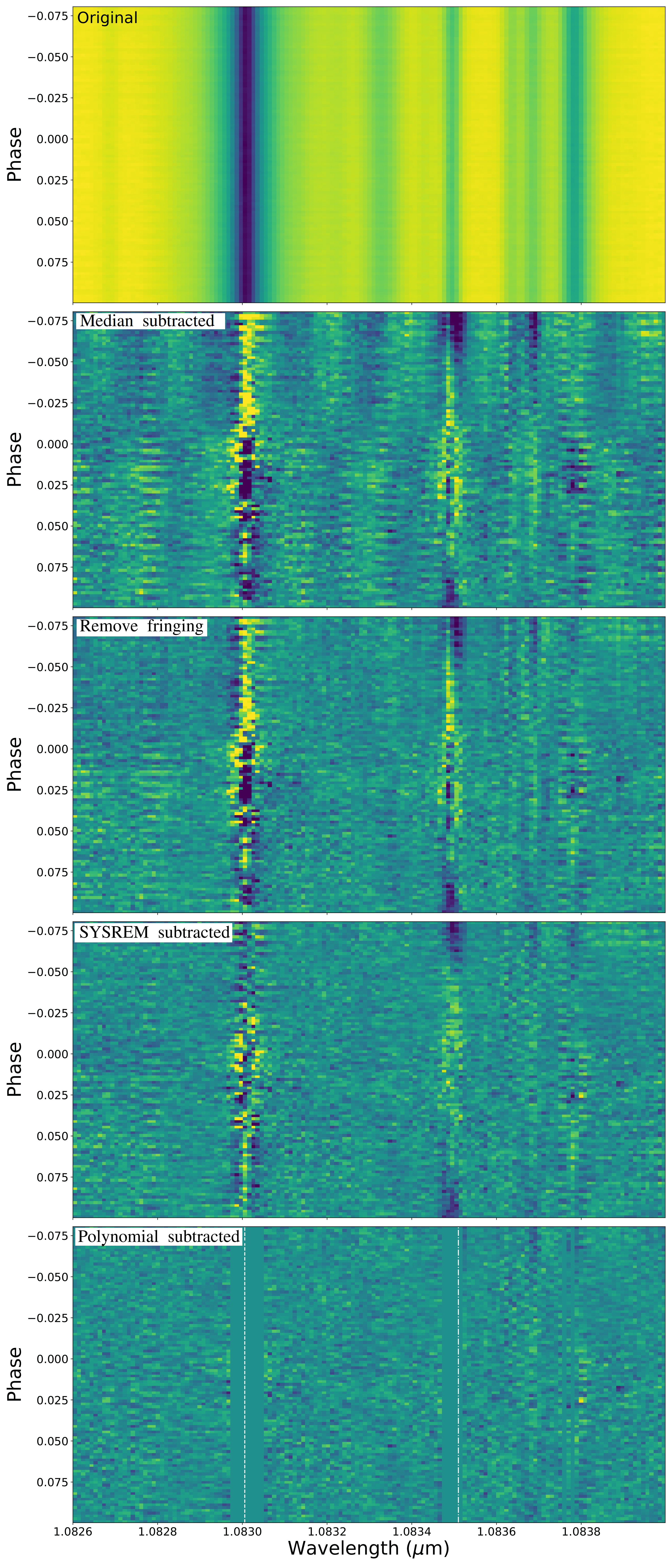}
    \caption{Pipeline steps: the original spectra; after removing fringing; after removing one SYSREM component; after masking variable lines and subtracting off continuum variations; after subtracting off continuum variations by fitting a polynomial to each spectrum.  In the last panel, the vertical white lines mark a stellar Si line at 1.0830057 \um (left) and a strong water line at 1.08351 \um (right).}
\label{fig:pipeline}
\end{figure}

We show the residuals image through various steps of the pipeline in Figure \ref{fig:pipeline} (top).  The most striking feature in the median subtracted image are the vertical bars caused by fringing, which are spaced 20 pixels apart.  A fast Fourier transform (FFT) of each spectrum reveals a prominent peak at a frequency of 0.052 pixel$^{-1}$.  To remove the fringing, we use \texttt{scipy} to apply a second order Infinite Impulse Response (IIR) notch digital filter with a frequency of 0.052 pixel$^{-1}$ and a quality factor of 15.  We apply the filter twice, which is sufficient to suppress the peak in the FFT without bringing the spectral power substantially below that of neighboring frequencies.  The subsequent panel in Figure \ref{fig:pipeline} shows the residuals image after fringing correction, demonstrating that the notch filter has effectively removed the fringing.

The most prominent features in the fringing-corrected residuals image (Figure \ref{fig:pipeline}) are the strongly variable columns.  These are deep telluric and stellar lines that vary for a variety of reasons, including the suboptimal performance of optimal extraction in deep lines (discussed in Figure \ref{subsec:opt_extract}), the inaccuracy of interpolation across deep lines, and the time-variable telluric absorption.  The line immediately to the right of center in the residuals image, for example, is a water absorption line at 1.08351 \um.  Its dependence on airmass is clear: as the night progresses, 55 Cnc first rises, then sets.  The line is dark at the beginning, brightens until the airmass reaches its minimum, and dims again.  The other strong line is a Si I line at 1.0830057 \um.

We correct for variability in telluric absorption using SYSREM \citep{mazeh_2007}.  SYSREM is a generalization of Principal Component Analysis that takes into account the errors on the data.  Like PCA, it identifies eigenvectors and eigenvalues which, when linearly combined, best explain the residuals image.  When we apply SYSREM to our residuals image, the first component has eigenvalues that closely track the airmass, showing that the algorithm is successfully identifying telluric variability.

\begin{figure*}[ht]
  \centering \includegraphics
    [width=\textwidth]{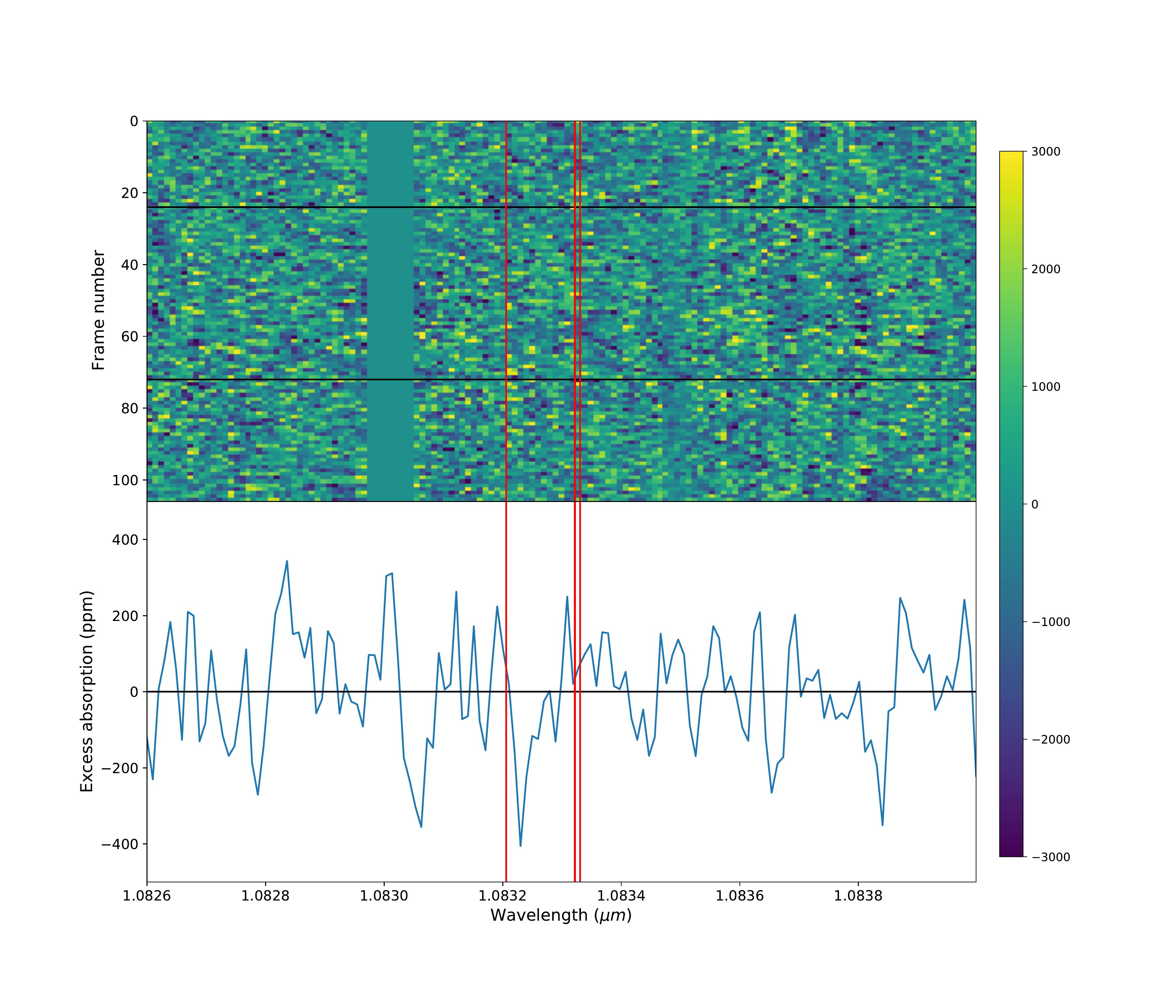}
    \caption{Top: Combined residuals image for both nights, showing excess absorption (ppm) in the stellar rest frame after all pipeline steps except shifting into the planetary rest frame.  The top and bottom horizontal black lines mark the beginning and end of transit, respectively.  The vertical red lines mark the helium triplet positions.  No helium absorption is evident.  Bottom: excess absorption during transit ($1 - F_{in}/F_{out}$) computed from the final residuals image for both nights.}
\label{fig:residuals_final}
\end{figure*}

After subtracting the first principal component identified by SYSREM, we compute the standard deviation of each column and mask the most variable columns.  This removes the prominent telluric line and all 4 of the most prominent stellar lines.  One iteration of SYSREM is not sufficient to remove the continuum variation seen in Figure \ref{fig:pipeline}.  We therefore remove it by fitting a third order polynomial with respect to column number for each row and subtracting the polynomial.  The residuals image, averaged across both nights, is shown in the bottom-most panel of Figure \ref{fig:pipeline}.  In Figure \ref{fig:residuals_final}, we additionally show a plot of excess absorption in the stellar frame.  There is no sign of helium absorption in either plot.

\subsection{Shifting into the planetary rest frame}
55 Cnc has an extremely high orbital speed of 230 km/s, causing it to accelerate by 130 km/s over the duration of the transit.  At every epoch, we use the barycentric Julian date to compute the radial velocity of the planet relative to the star, shift the residual spectrum accordingly, and resample onto the common wavelength grid using linear interpolation.  The covariances introduced by linear interpolation are properly computed and propagated.  We then take the mean of all the shifted in-transit residual spectra to arrive at the final excess absorption spectrum.  The average spectrum across both nights is taken to be the fiducial excess absorption spectrum.

\begin{figure*}[ht]
\subfigure {\includegraphics
    [width=\textwidth]{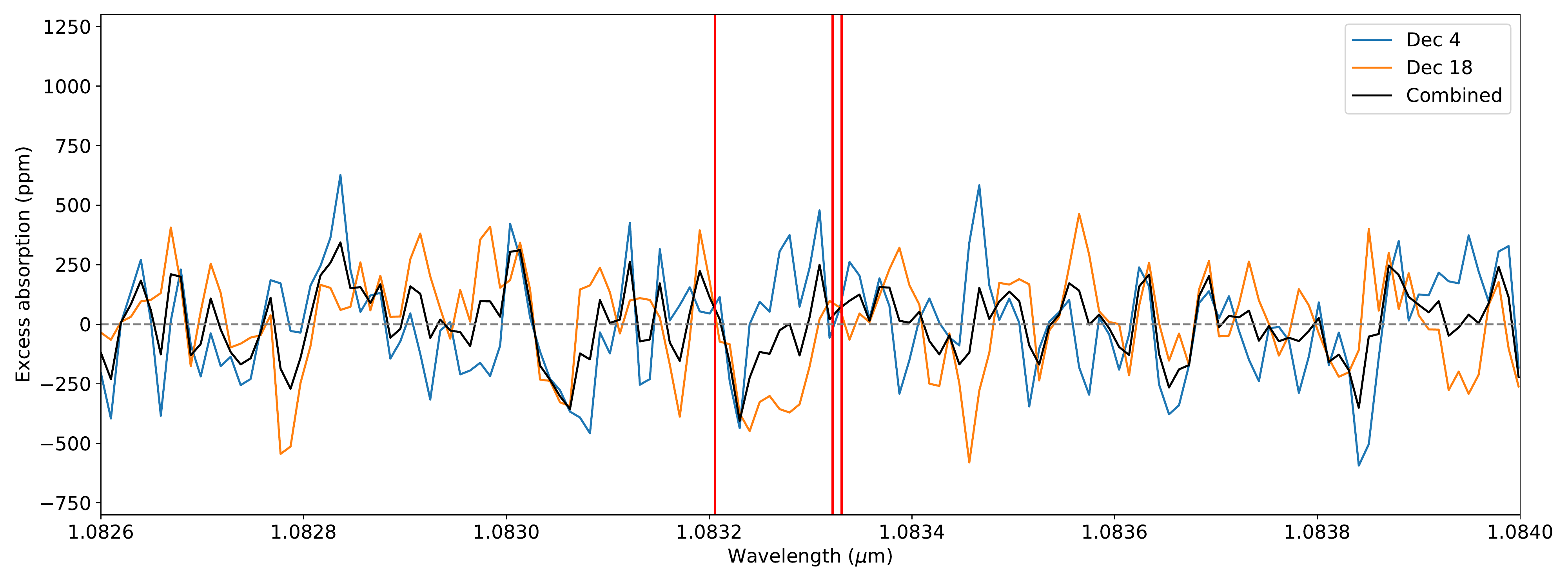}}\qquad\subfigure {\includegraphics
    [width=\textwidth]{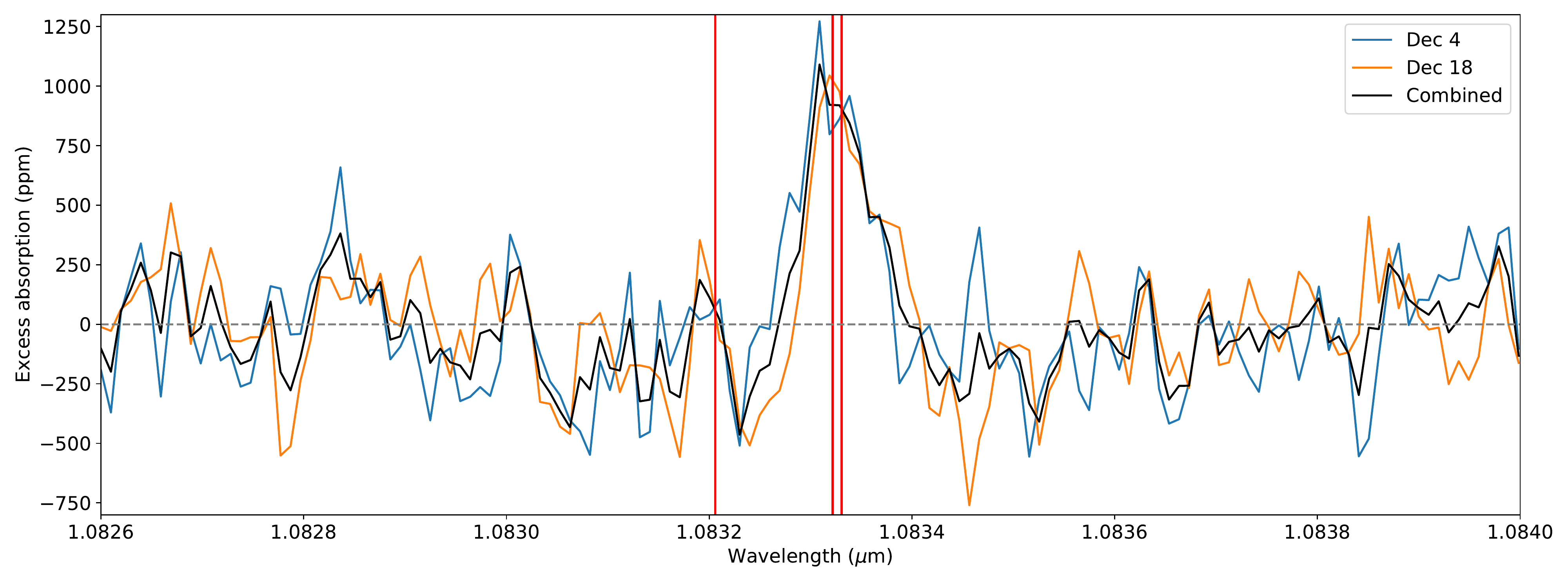}}
    \caption{Top: excess absorption during the transit in the planet frame as a function of wavelength, averaged over both nights.  Helium absorption from the planet would result in a positive signal at the position of the red vertical lines.  Due to the nature of ground-based high resolution spectra, the white light transit cannot be detected.  Bottom: same as above, except an absorption signal was injected into the spectra before any reduction was undertaken.  The injected signal has a maximum excess absorption of 1300 ppm, corresponding to the 2.5D hydrodynamic model with the weakest absorption (dashed blue curve in Figure \ref{fig:simulation-spec}).}
\label{fig:planet_frame_excess_absorption}
\end{figure*}

The excess absorption from both nights, as well as the averaged excess absorption, is shown in Figure \ref{fig:planet_frame_excess_absorption}.  The combined spectrum has a standard deviation of 146 ppm, smaller than the planet's white light transit depth of 350 ppm, although some amount of correlated noise is present.  No sign of helium absorption can be seen.

The bottom panel of Figure \ref{fig:planet_frame_excess_absorption} shows what the excess absorption spectrum would look like with a 1300 ppm helium absorption signature.  The absorption profile was taken from the 2.5D hydrodynamic model with the weakest absorption (solid green curve in Figure \ref{fig:simulation-spec}).  This figure shows that the reduction process subtracts out at most 20\% of the signal, and that even the 2.5D hydrodynamic model with the weakest absorption is ruled out by our data.

\subsection{Constraints on absorption}
To quantify the constraint on excess absorption that our data provide, we used the nested sampling code \texttt{dynesty} \citep{speagle_2019}.  We modelled the data as a scaled and vertically shifted version of the 2.5D hydrodynamic model with the weakest absorption (solid green curve in Figure \ref{fig:simulation-spec}, corresponding to a H-dominated atmosphere with $10^{-10}$ envelope fraction and 1 year dispersal timescale).  The observed excess absorption spectrum (Fig. \ref{fig:planet_frame_excess_absorption}) is truncated to 10,831--10,835 \ang to avoid the need to model any unsubtracted low-frequency variations in the data.  Instead, the low-frequency variations are accounted for by vertically shifting the model to match the data.

We compute a covariance matrix for the excess absorption spectrum using the errors from the optimal extraction algorithm propagated forward through the pipeline.  This matrix accounts for the covariances caused by the two linear interpolations--first onto a common wavelength grid, and second into the planetary rest frame.   To account for unmodelled systematics, we multiply the covariance matrix by a free parameter $e^2$, where $e > 1$.   The log likelihood is given by:

\begin{align}
    \ln{L} = -\frac{1}{2} r^TK^{-1}r - \frac{1}{2}|K| - \frac{N}{2}\ln{2\pi},
\end{align}

\noindent where $K$ is the covariance matrix and $r$ is the residuals. 

\begin{figure}[ht]
  \centering \includegraphics
    [width=0.5\textwidth]{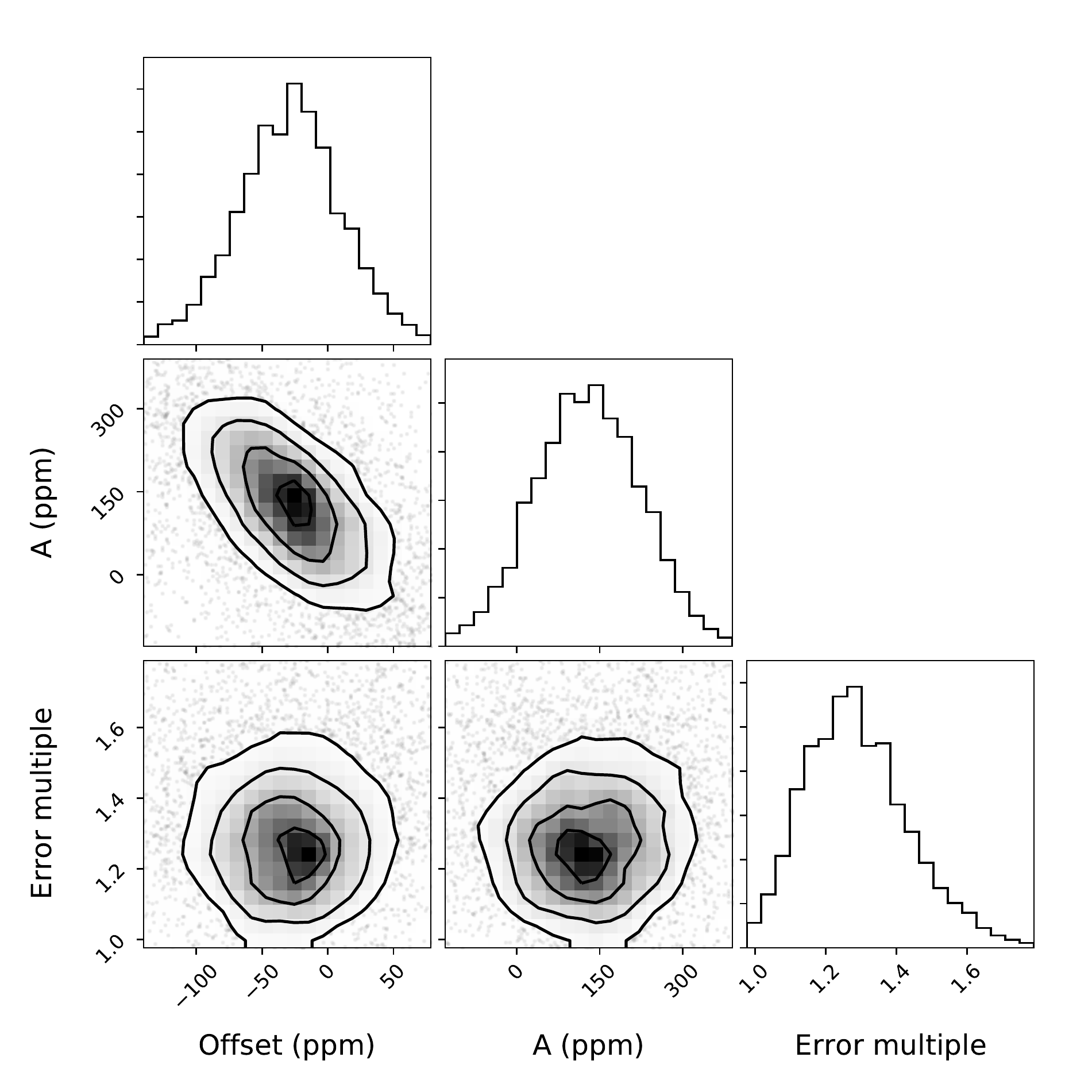}
    \caption{2D posteriors from our fit to the data, constraining the maximum excess absorption (A).  A is 1.3$\sigma$ from 0.}
\label{fig:nested_constraints}
\end{figure}

Figure \ref{fig:nested_constraints} shows the results of our nested sampling run.  The free parameters are the vertical offset of the model, the maximum excess absorption ($A$), and the error multiple $e$ whose square multiplies the covariance matrix.  Our data is consistent with zero excess absorption, and the best-fit error inflation parameter $e$ indicates that our errors are likely underestimated by approximately 25\%.  The preferred vertical offset is consistent with zero, as expected from visual inspection.
Our fit places a 90\% upper limit on $A$ of 250 ppm, corresponding to an equivalent width of 0.27 m\ang.  To see whether $A$ is significantly non-zero, we performed a nested sampling run with the amplitude removed as a free parameter.  The resulting log Bayesian evidence log(Z) was 1.3 higher in this run, indicating that the data prefers a model with no absorption.

\begin{figure}[ht]
  \includegraphics
    [width=0.5\textwidth]{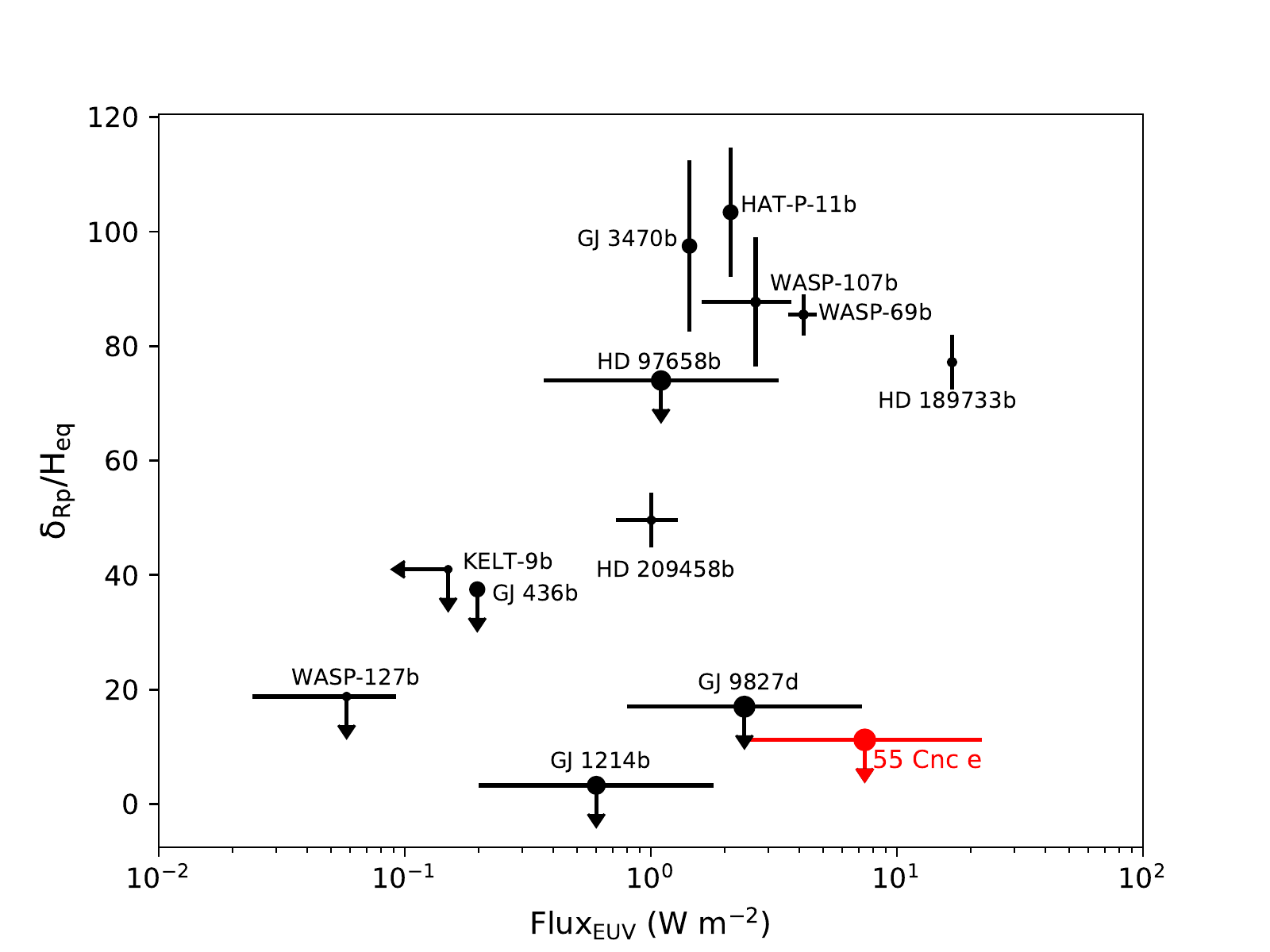}
    \caption{55 Cnc e in context, with the size of each point inversely proportional to the planetary radius.  The y-axis shows $\delta_{Rp}/H_{eq}$, the ratio of the increase in apparent radius at 1083 nm to the scale height at equilibrium temperature.  The only super Earth with a helium measurement or non-detection, 55 Cnc e is highly irradiated and has a very tight observational constraint compared to its peers. Data for other planets was taken from \citealt{kasper_2020}.  We did not include the helium non-detections of AU Mic b \citep{hirano_2020}, WASP-52b \citep{vissapragada_2020}, K2-100b \citep{gaidos_2020a} or K2-25b \citep{gaidos_2020b}, which are less sensitive (the upper limits on $\delta_{Rp}/H_{eq}$ are all greater than 100).}
\label{fig:context}
\end{figure}

The constraint on $A$ can also be expressed in scale heights, where the scale height $H_{eq}=\frac{k_BT_{eq}}{\mu g}$ is computed assuming an equilibrium temperature of 2000 K and a hydrogen-dominated atmosphere.  This metric was first proposed by \cite{nortmann_2018}.  We obtain an upper limit on the increase in apparent planet radius at 1083 nm: $\delta_{Rp}/H_{eq} < 11$ with $H_{eq}=350$ km, which we put into context in Figure \ref{fig:context} by plotting the EUV flux and $\delta_{Rp}/H_{eq}$ of all planets with helium detections or non-detections.  The EUV flux experienced by 55 Cnc e is obtained from \cite{salz_2016}, who calculate $F_{EUV} = 7.4$ W/m$^2$ by applying the scaling relation of \cite{linsky_2014} to the Lyman alpha luminosity.  Following \cite{kasper_2020}, we conservatively adopt 3x error bars on the EUV flux on either end, as different EUV reconstruction techniques give results that are discrepant by an order of magnitude in some cases (i.e. \citealt{salz_2015b, drake_2020}).  Figure \ref{fig:context} shows that, relative to other planets with helium measurements, 55 Cnc e is a highly EUV-irradiated planet with a tight upper limit that is far below the $\delta_{Rp}/H_{eq}$ of successful detections for other planets.  

\section{Modeling}
We next turn to helium outflow models to explore what our non-detection of helium absorption might mean for 55 Cne e's atmospheric composition and corresponding mass loss rate.  We start by performing order-of-magnitude estimates of the mass loss rate from analytical formulae.
Next, we use three independent models to interpret the observational result.  First, an isothermal Parker wind model \citep{oklopcic_2018} puts constraints on the 2D parameter space of temperature and mass loss rate, but does not constrain either parameter independently.  Our second model, The PLUTO-\cloudy Interface (TPCI), can model the outflow in a 1D fashion given the stellar XUV spectrum and predict the mass loss rate, temperature profile, and absorption spectrum.   Our third and most sophisticated model is a 2.5D model \citep{wang_2018} which combines ray-tracing radiative transfer, real-time non-equilibrium thermochemistry, and hydrodynamics to model the outflow, assuming it is symmetric about the star-planet axis.  This model can also predict the mass loss rate, temperature profile, and absorption spectrum, given an assumed atmospheric composition and envelope fraction.

We compare the TPCI and 2.5D predictions directly to observations.  Since the Parker wind model requires both the mass loss rate and the exosphere temperature as input parameters, we take a typical exosphere temperature from the TPCI model as a reasonable estimate, and use it within the framework of the Parker wind model to constrain the mass loss rate.

For convenience, we compile the take-away results of the models we considered in Table \ref{table:models_summary}.  In many of these models, we consider a range of possible parameters and atmospheric compositions.  The range of values we give in the table reflect the range of parameters and compositions.

\begin{table*}[ht]
  \centering
  \caption{Summary of Model Predictions}
  \begin{tabular}{c c c c}
  \hline
  	  Model & Dimensions & $\dot M$ (g/s) & Peak Absorption (ppm) \\
      \hline
      Energy-limited & 0 & $>$2\e{9} & N/A \\
      Semi-empirical & 0 & 8\e{9}--1\e{11} & N/A\\
      Parker wind & 1 & N/A & 900-1800$^*$  \\
      TPCI & 1 & 0.75--1.1\e{10} & 800--1000 \\
      2.5D & 2.5 & 0.70--1.5\e{10} & 1400--2400 \\
      \hline
  \end{tabular}
  \label{table:models_summary}
  \tablecomments{$^*$Assumes the mass loss rate and exosphere temperature predicted by TPCI}
\end{table*}

\subsection{Analytic estimates of escape rate}
Before running complex hydrodynamic simulations, it is useful to perform rough analytical estimates of the escape rate to illustrate the typical magnitudes involved and their dependence on stellar and planetary quantities.  The simplest way to estimate the escape rate is to assume that it is limited by the X-ray and extreme UV stellar radiation hitting the planet.  The energy per unit mass required to escape the planet's gravity well is $dE/dm = -GM_p/R_{XUV}$, while the rate at which XUV radiation impinges upon the planet is $dE/dt = \frac{L_{XUV}}{4\pi a^2} \pi R_p^2$.  Assuming a fraction $\eta$ of the energy goes toward driving mass loss, we obtain:

\begin{align}
    \frac{dm}{dt} = \frac{\eta}{4} \frac{R_p^3 L_{XUV}}{GM_p a^2}
\end{align}

$R_p$ is the radius of the XUV photosphere, but we take it to be the optical transit radius in order to obtain a conservatively low mass loss rate. We obtain $L_{XUV}=10^{27.70}$ erg/s from \cite{salz_2016}, who in turn obtain it by applying the scaling law in \cite{linsky_2014} to the star's Lyman alpha luminosity.  We assume a low efficiency $\eta=0.15$, in line with Monte Carlo heating models \citep{shematovich_2014,ionov_2015}, and obtain 2\e{9} g/s.  Because the value of $\eta$ and $L_{XUV}$ are both highly uncertain, this should be regarded as an order-of-magnitude estimate only.

A slightly more sophisticated approach to estimating the mass loss rate is to adopt the semi-empirical expressions of \cite{wang_2018}, who derived their expression from 2.5D numerical simulations of mass loss.  Their equation is:

\begin{align}
\begin{split}
    \dot M = 4.5 \times 10^{-10} M_{\earth}/yr \cdot (\frac{R_{EUV}}{5R_{\Earth}})^2 \\\text{max}(F', F'^{0.6}) \text{min}(1, M'^{-0.5})\\
\end{split}    
\end{align}
\begin{align}
    M' &= \frac{M_c}{10 M_{\Earth}}\\
    F' &= \frac{F_{EUV}}{5 \times 10^{14} erg/s/\textrm{\mathang}}
\end{align}

For 55 Cnc, the X-ray luminosity is only 10\% of the total XUV luminosity, so the terms EUV and XUV are more or less interchangeable.  We use three methods to determine $R_{EUV}$, the apparent radius of the planet in the EUV.  First, we use the radius implied by the optical transit depth: 1.9$R_\Earth$.  This radius leads to a mass loss rate of 8\e{9} g/s.  Second, we use the Hill radius, $R_{Hill} = a (\frac{M_p}{3M_s})^{1/3} = 7.5 R_\Earth$.  This leads to a mass loss rate of 1\e{11}  g/s.  These two mass loss rates are the lower and upper limits of what is reasonable under the semi-empirical framework of \cite{wang_2018}.

The third method for estimating $R_{EUV}$ is to follow \cite{wang_2018} in assuming that the EUV photosphere is at $\rho=10^{-13}$ g cm\textsuperscript{-3}.  To find the radius that corresponds to this density, we assume the atmosphere is isothermal, and need to find one point with known r and $\rho$.  \cite{wang_2018} (following \citealt{owen_2012}) pick the radiative-convective boundary (RCB), and use a parameterized opacity to calculate the position of the RCB.  Unfortunately, this calculation involves many unknown quantities, such as the Kelvin-Helmhotz timescale and the envelope fraction.  It may also be particularly ill-suited to thin atmospheres, where the radiative-convective boundary may be very close to the surface and at a much higher temperature than $T_{eq}$.  We therefore assume that P=100 mbar corresponds to the radius inferred from the optical transit depth, 1.9 $R_\Earth$.  100 mbar is the approximate photosphere of transit observations in the optical for solar metallicity planets with thick atmospheres, making this guess more accurate than trying to estimate the RCB radius and pressure.

Assuming r(P=100 mbar) = 1.9 $R_\Earth$, we obtain $\rho_{phot} = \frac{P \mu}{k_B T_{eq}} = 1.4 \times 10^{-6}$ g/cm$^3$ and $\beta \equiv \frac{GM_c \mu}{R_p k_B T_{eq}} = 37$, and can calculate $R_{EUV}$:

\begin{align}
    R_{EUV} &= \frac{R_{phot}}{1 + \beta^{-1} \ln{(\rho_{EUV} / \rho_{phot})}}\\
            &= 3.4 R_\Earth
\end{align}

This implies $\dot M$ = 3\e{10} g/s.

\subsection{1D Parker wind model}
Now that we have a rough sense for the expected mass loss rate for a hydrogen- and helium-rich atmosphere, we can ask whether mass loss rates in this range are definitively ruled out by our helium non-detection.  We use the Parker wind model of  \cite{oklopcic_2018} to translate our helium non-detection into a joint constrain on the mass loss rate and the exosphere temperature.  This study treats the outflow as an isothermal, radially symmetric Parker wind from a hydrogen-dominated atmosphere.  The model calculates the population levels of singlet and triplet states by balancing recombination, photoionization, collisional (de)-excitation, and radiative decay as a function of altitude.  Using the helium number density and the population level of the triplet state as a function of altitude, it then calculates the transit depth as a function of wavelength.  

We simulate absorption spectra for a 2D grid of mass loss rates ranging from $10^7$ to $10^{11}$ g/s, and exospheric temperatures ranging from 2500 to 9000 K, in order to better quantify the limits our data place on the planet's present-day mass loss rate.  For each combination of mass loss rate and exosphere temperature, we compute an excess absorption spectrum and calculate the corresponding log likelihood of the observational data given the model.  We then compute $\Delta$ln(L): the difference between the ln(L) of model versus that of a zero absorption model.

\begin{figure}[ht]
  \includegraphics
    [width=0.5\textwidth]{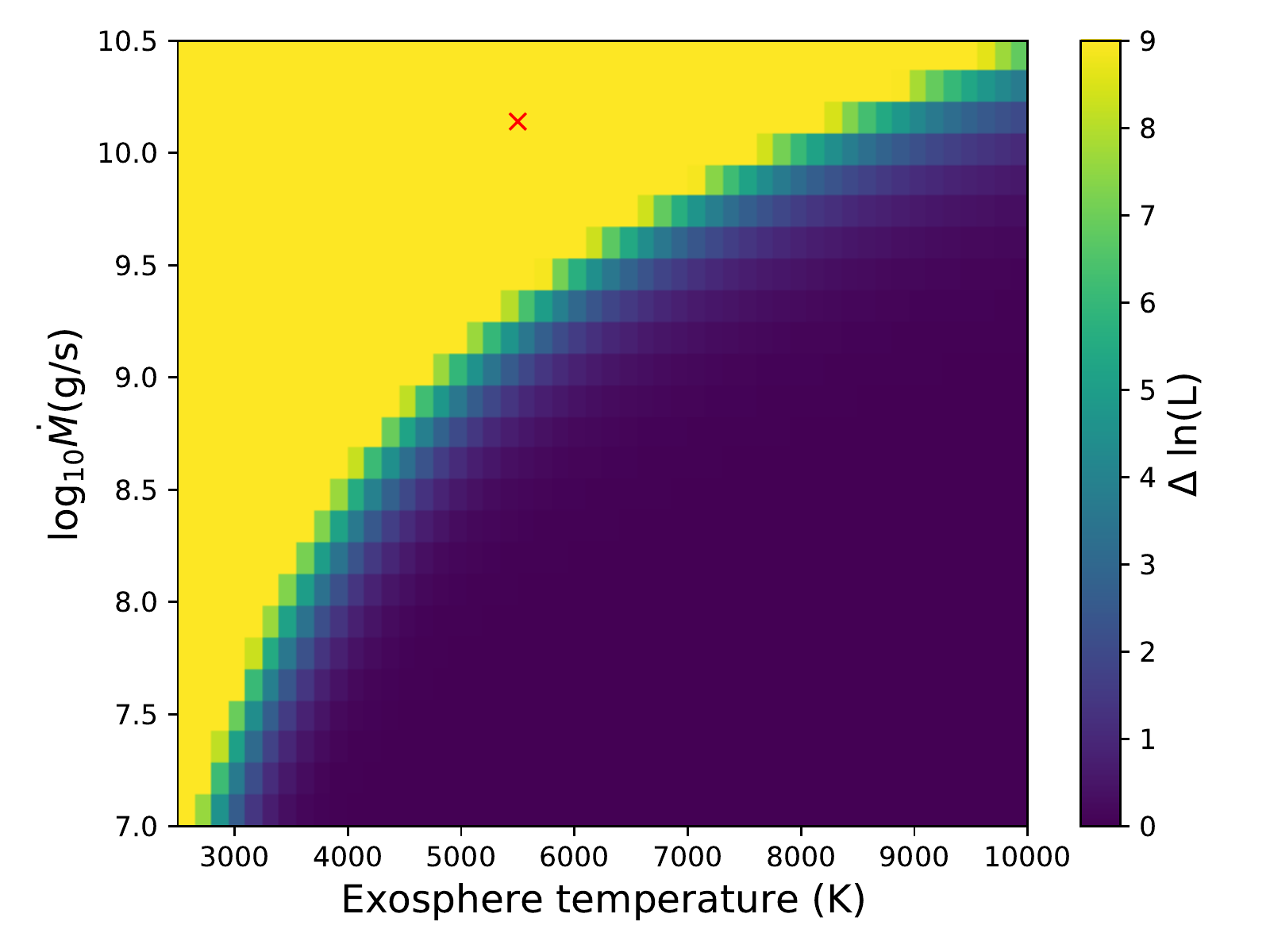}
    \caption{The confidence with which each combination of mass loss rate and exosphere temperature is ruled out, according to the 1D Parker wind model of \cite{oklopcic_2018}.  The red x indicates the mass loss rate and approximate exosphere temperature predicted by the non-isothermal \cite{salz_2016} model.}
\label{fig:mass_loss_constraint}
\end{figure}

\begin{figure}[ht]
  \includegraphics
    [width=0.5\textwidth]{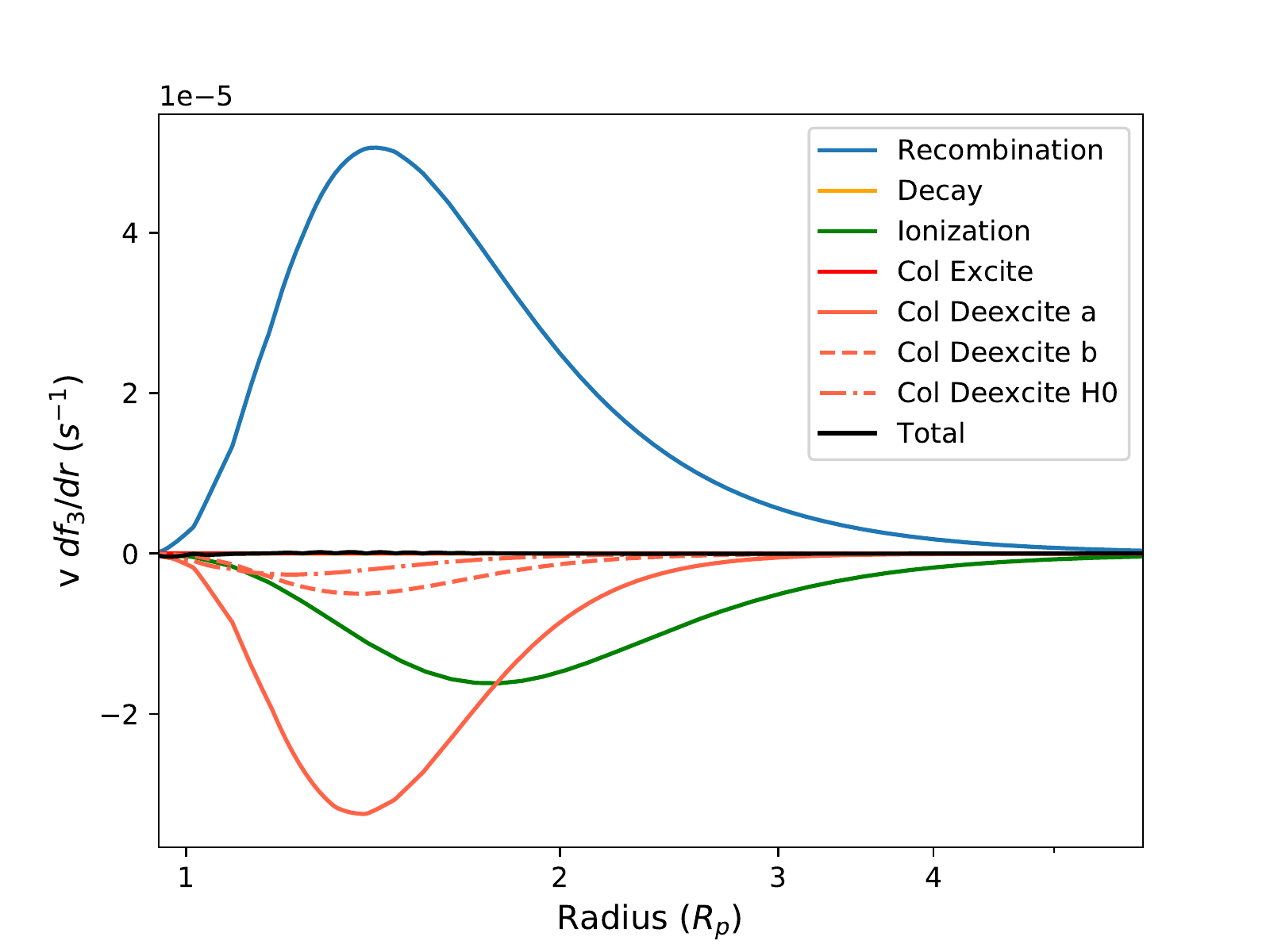}
    \caption{Production and destruction rates of triplet helium due to various processes.  From a Parker wind model of 55 Cnc e with a mass loss rate of 1.38\e{10} g/s and exosphere temperature of 5000 K.}
\label{fig:parker_rates}
\end{figure}

We show the resulting $\Delta ln(L)(\dot M, T_0)$ in Figure \ref{fig:mass_loss_constraint}.  In these models higher temperature exospheres have faster outflow velocities and correspondingly weaker helium absorption for a given mass loss rate. At the planetary surface, for example, a 5000 K exosphere with a mass loss rate of $1.4\times10^{10}$ g/s requires a 0.16 km/s wind; a 10,000 K exosphere with the same mass loss rate requires a 3 km/s wind.  A faster wind implies lower density at the same mass loss rate, decreasing helium absorption.  This is both because there are fewer helium atoms per cubic volume, and because the fraction of helium atoms in the triplet state is lower.  The latter, in turn, is because the triplet state is primarily populated by recombination (see Figure \ref{fig:parker_rates}), and a lower density means a lower recombination rate.  For these reasons, the mass loss constraint is less stringent at higher temperatures.  

The Parker wind model does not provide a way to estimate the exosphere temperature, which makes it impossible to constrain the mass loss rate without the help of another model.  The TPCI model predicts a peak exosphere temperature of 5000-6000 K (\citealt{salz_2016}; see next subsection for details).  At $T_0=5000$ K, the mass loss rate is constrained to $\dot M < 10^{9.1}$ g/s ($\Delta ln(L) < 7$), or $\dot M < 10^{8.8}$ g/s ($\Delta ln(L) < 3$).  At $T_0=6000$ K, the mass loss rate is constrained to $\dot M < 10^{9.5}$ g/s ($\Delta ln(L) < 7$), or to $\dot M < 10^{9.2}$ g/s ($\Delta ln(L) < 3$), where $\Delta ln(L)$ thresholds of 7 and 3 correspond to likelihood ratios of 1100 and 20, respectively.  In summary, the Parker wind model gives an upper limit on the mass loss rate of $\sim10^9$ g/s for exospheric temperatures predicted by TPCI.

Lastly, we use this model to gain intuition on the physical processes that determine the triplet helium fraction, and therefore the helium absorption strength.  Figure \ref{fig:parker_rates} plots, as a function of radius, the production and destruction rates of triplet helium due to the processes considered by \cite{oklopcic_2018}: recombination, radiative decay, ionization, collisional excitation/deexcitation with electrons, and collisional deexcitation with neutral hydrogen atoms.  Recombination is the most important production mechanism, while destruction is due to a combination of collisional deexcitation with electrons and ionization: the former is dominant close to the planet while the latter is dominant far from the planet, as one would expect.  Collisional excitation and radiative decay are negligible.  The production and destruction rates very nearly cancel out, indicating that the triplet helium fraction is mostly in equilibrium and that advection is not significant.

\subsection{1D PLUTO-\cloudy hydrodynamic model}
\begin{figure*}[ht]
  \includegraphics
    [width=\textwidth]{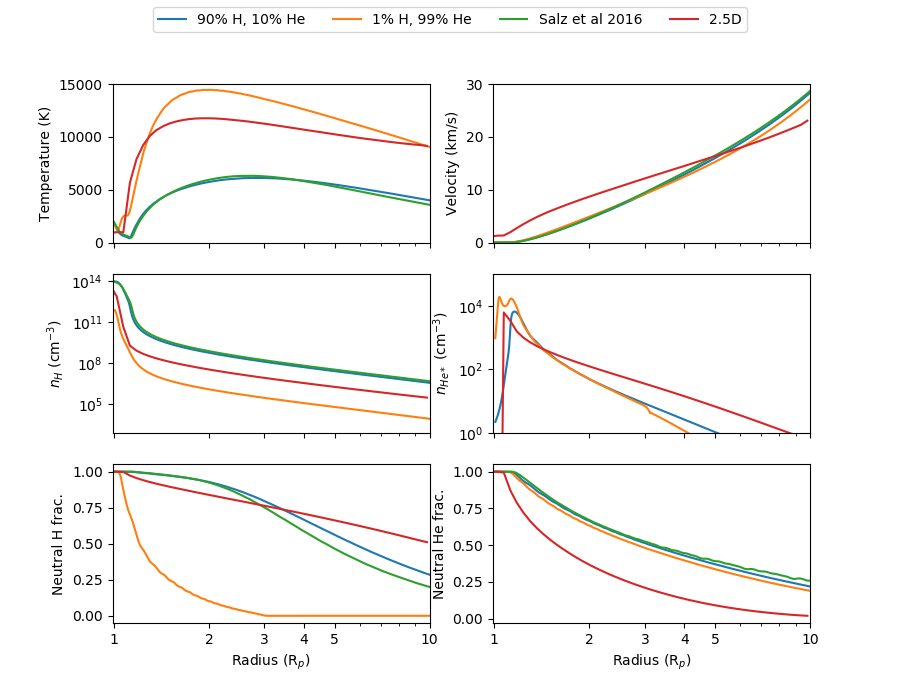}
    \caption{Profiles of various physical quantities for the hydrogen- and helium- dominated TPCI models in addition to the 2.5D model.  The low-amplitude oscillations in the helium-dominated profiles are numerical artifacts resulting from the non-zero advection length.  In green are the results from \citealt{salz_2016}, who compute all quantities plotted here except for the helium triplet density.}
\label{fig:tpci_models_comp}
\end{figure*}

\begin{figure}[ht]
  \centering \subfigure {\includegraphics
    [width=0.5\textwidth]{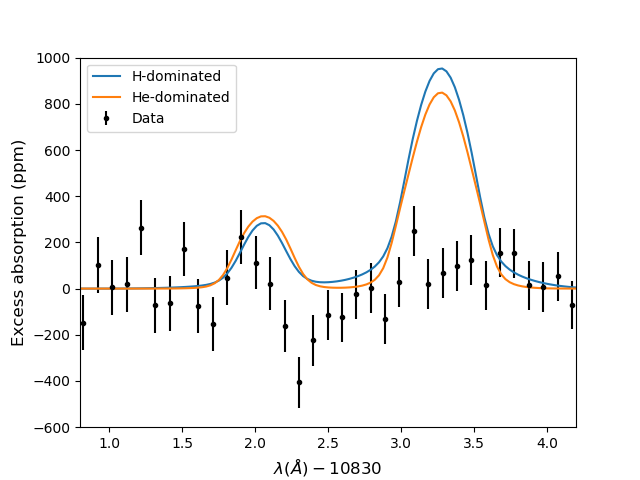}}\qquad
  \subfigure {\includegraphics
    [width=0.5\textwidth]{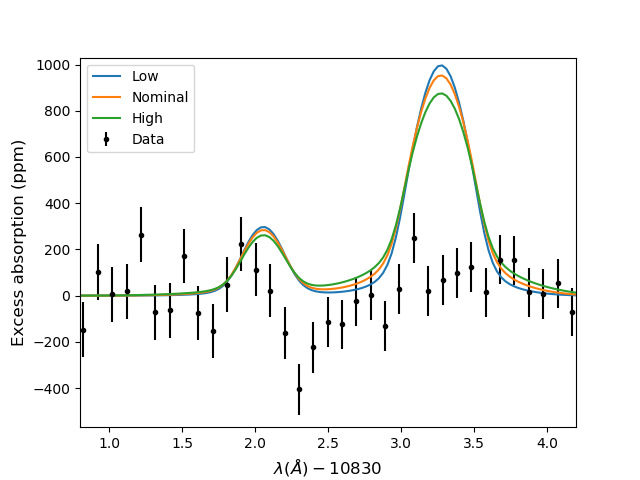}}
    \caption{Top: TPCI hydrogen-rich (90\% H, 10\% He) and helium-rich (1\% H, 99\% He) models are both ruled out by our observational data.  Bottom: increasing or decreasing the stellar XUV flux by a factor of 2 does not significantly change the absorption signal in the TPCI hydrogen-rich model.}
\label{fig:tpci_obs_comparison}
\end{figure}

In this section we use a 1D, spherically symmetric radiative-hydrodynamical simulation to predict the exospheric temperature structure, mass loss rate, and metastable helium absorption signal for 55 Cnc e under several different scenarios.  
\cite{salz_2016} previously used this model to simulate hydrogen-rich exospheres for several planets, including 55 Cnc e, but did not calculate the predicted absorption signal from metastable helium.  The \cite{salz_2016} model predicts an exospheric temperature of 3000-6000 K and an outflow velocity at large distances of 10-15 km/s.  It predicts a mass loss rate of 1.4\e{10} g/s, corresponding to 0.9\% of the planetary mass per Gyr.  As the paper shows, the predicted Lyman-$\alpha$ signal is consistent with the non-detection by \cite{ehrenreich_2012}.  This mass loss rate would imply that 55 Cnc e started life as a sub Neptune with a hydrogen/helium envelope of $\sim$10\% by mass. 

We use TPCI to reconstruct the \citealt{salz_2016} model and calculate the corresponding helium absorption signal.  This also allows us to explore other compositions, including a helium-dominated atmosphere.

This model couples PLUTO, a hydrodynamics code that can work in 1, 2, or 3 dimensions \citep{mignone_2007}, and \cloudy \citep{ferland_2013}, a 1D plasma simulation and spectral synthesis code.  PLUTO and \cloudy are linked in The PLUTO-\cloudy Interface \citep[TPCI;][]{salz_2015}.  In TPCI, \cloudy calculates the equilibrium chemistry and ionization state of the medium given a radiation field, computes the heating and cooling rates of the new state, and feeds this information to PLUTO.  PLUTO heats or cools the medium appropriately, evolves the medium hydrodynamically, and provides the new state to \cloudy, after which the cycle restarts.  Both PLUTO and \cloudy are sophisticated, publicly available, and general-purpose codes that have been applied to a variety of astrophysical problems, ranging from exoplanet mass loss to the magnetic fields of neutron stars (PLUTO; e.g. \citealt{sur_2020}) to high-redshift gamma ray bursts (\cloudy; e.g. \citealt{shaw_2020}).  \cite{salz_2016} adopt a 1D spherically symmetric model, and do not include the planetary magnetic field.  \cloudy includes the 30 lightest elements, from hydrogen to zinc, and accounts for many physical processes, including radiative and collisional ionization/recombination, inner shell ionization, and charge exchange.  However, \cite{salz_2016} assume a purely atomic hydrogen and helium atmosphere.

As input to the PLUTO-\cloudy model, the authors use the X-ray luminosity measured by XMM-Newton in April 2009, namely 4.6\e{26} erg/s between 5--100 \ang \citep{sanz-forcada_2011}.  This is 2.4 times lower than the flux measured by Chandra, which was obtained simultaneously with the \emph{HST} Lyman alpha observations in March/April 2012 \citep{ehrenreich_2012}.  This is likely because March/April 2012 was right at the minimum of the $10.5 \pm 0.5$ year stellar activity cycle \citep{bourrier_2018}, while April 2009 was 3 years from minimum and 2.3 years from maximum.  Because our December 2019 observations were almost exactly one stellar cycle after April 2009, we expect the star's X-ray properties to be similar to those observed by \cite{sanz-forcada_2011} and adopted by \cite{salz_2018}, and more favorable for mass loss measurements than the conditions encountered by \cite{ehrenreich_2012}.

Because we do not know exactly what settings or what version of TPCI they used we are unable to replicate their results exactly, but we match their temperature profile to better than 500 K and their mass loss rate to within 25\% accuracy (see Figure \ref{fig:tpci_models_comp}).  This is far smaller than the factor of several uncertainty they report as the inherent model error (their Table 2), resulting from factors such as the neglected 3D structure (4x uncertainty), uncertain irradiation strength (3x), and neglect of magnetic fields (2x).  Following Salz et al., we neglect molecules and elements other than hydrogen and helium in our simulation.  

We find that the advection length--roughly speaking, the resolution of \cloudy's advection calculations-- is a crucial parameter for these models.  Smaller values lead to more accurate results, but take longer to converge.  Large values lead to spurious spatial oscillations in the temperature and ionization state.  We adopt an advection length of 0.15 planetary radii, which we find is small enough that the oscillations in temperature are of order 0.2\% for the helium-dominated run and 0.06\% for the hydrogen-dominated run.  We run the simulations for 150 days of model time (roughly 1000 sound crossing times) while neglecting advection.  We then turn on advection, which slows down the run by a factor of $\sim$80, and let the simulation run for another 150 days of model time.  We monitor the evolution of the temperature, density, velocity, and mass loss rate (calculated as $4\pi r^2 \rho(r) v(r)$) profiles to verify that the simulation has in fact converged, with temperatures fluctuating by less than 50 K and the mass loss rate by less than 0.1\%.

After reproducing the hydrogen-dominated model from \cite{salz_2016}, we run a second TPCI simulation for a helium-dominated atmosphere with 99\% He and 1\% H by number to explore a scenario where slow mass loss over many Gyr fractionated the atmosphere.  

Since \cloudy computes the level populations of every species, we configure it to report the number density of helium atoms in the triplet state at every radial coordinate.  We use this number density profile, in addition to the temperature and velocity profiles, to compute the excess absorption spectrum.

Figure \ref{fig:tpci_models_comp} compares the profiles for temperature, velocity, hydrogen density, helium triplet density, and neutral fraction for the hydrogen- and helium-dominated TPCI models.  Using these profiles, we computed the mass loss rate as $\dot{M} = 4\pi r^2 \rho v/4$.  The division by 4 follows \cite{salz_2016} and is meant to account for the 3D nature of the outflow, as we simulate only the sub-stellar point.  We obtain a mass loss rate of 1.1\e{10} g/s for the hydrogen-dominated scenario and 7.5\e{9} g/s for the helium-dominated scenario.  

In Figure \ref{fig:tpci_obs_comparison}, we compare the predicted excess absorption from our TPCI models to the observed excess absorption spectrum. Taking into account the interpolation-induced covariance between the data points, we find that a zero absorption model is preferred over the TPCI hydrogen-dominated model by $\Delta$ln(L) = 126, and preferred over the helium-dominated model by $\Delta$ln(L) = 100.  Although these estimates do not account for the error due to systematics or variable tellurics, it can be seen visually that the predicted absorption for both models is ruled out.

\begin{figure*}[ht]
  \includegraphics
    [width=\textwidth]{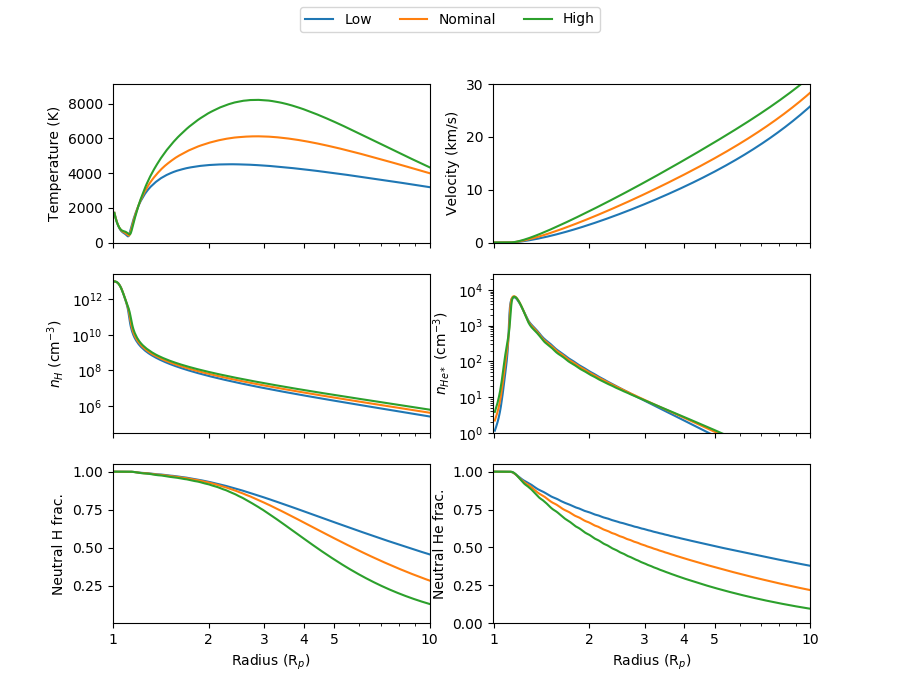}
    \caption{Profiles of various physical quantities for hydrogen-dominated TPCI models of different stellar XUV flux: nominal, 2x nominal (high), and half nominal (low).}
\label{fig:low_high_flux}
\end{figure*}

Critical to all photoevaporation models is the stellar EUV luminosity, which cannot be directly observed and must be reconstructed from X-ray and/or Lyman-$\alpha$ flux.  However, different EUV reconstruction techniques give results that are discrepant by an order of magnitude in the worst-case scenario (i.e. \citealt{salz_2015b, drake_2020}), although inactive main sequence stars like 55 Cnc e can be modelled more accurately.  To determine the effect that uncertainties in the EUV luminosity have on the helium absorption signal, we ran TPCI simulations for the hydrogen-dominated scenario at twice and half the nominal stellar flux.  The profiles of various physical quantities with respect to radius is shown in Figure \ref{fig:low_high_flux} for the low-flux, nominal, and high-flux scenarios, while the helium absorption signals are compared in Figure \ref{fig:tpci_obs_comparison}.

The high flux scenario has a mass loss rate 3 times higher than the low flux scenario, implying a slightly sub-linear scaling of mass loss rate with respect to flux.   Doubling the stellar flux substantially increases the exosphere temperature and outflow velocity while substantially decreasing the neutral fraction of both hydrogen and helium.  The hydrogen number density is moderately higher at higher flux, but the triplet helium number density is remarkably insensitive to flux.  This insensitivity means that the excess absorption is barely affected by uncertainties in the stellar flux--it is only 20\% lower in the high-flux scenario than in the low-flux scenario.

\subsection{2.5D hydrodynamic simulations}
\label{sec:simulations}

\begin{figure*}[ht]
  \includegraphics
    [width=\textwidth]{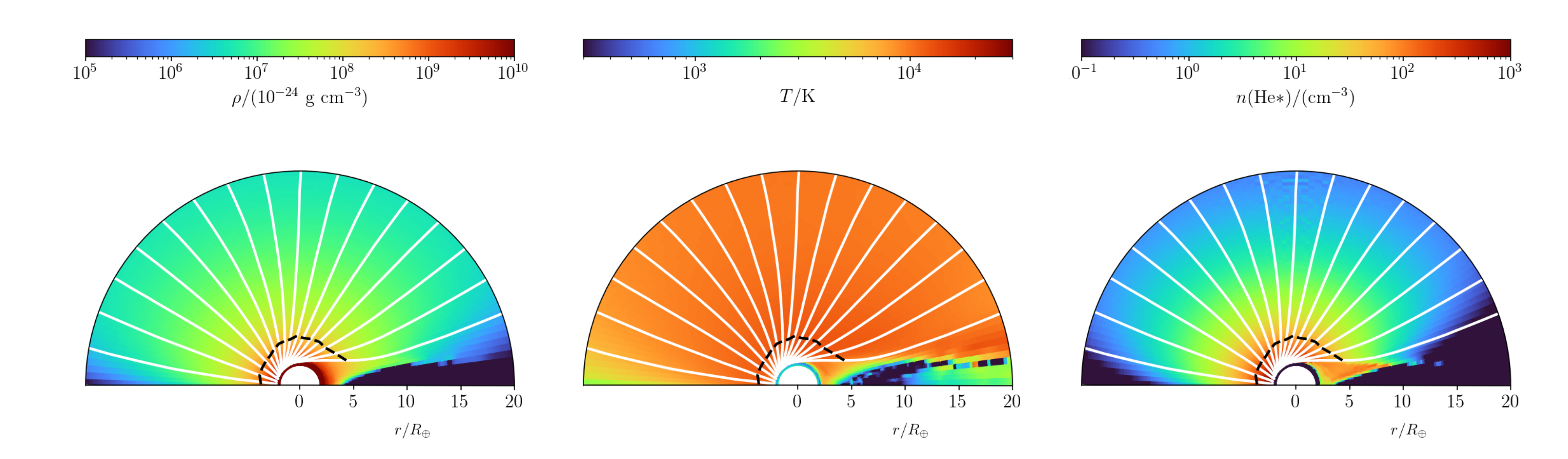}
    \caption{Density, temperature, and triplet helium density from the fiducial 2.5D model.  The star is toward the left, and the simulation volume is represented by revolving the semicircles about their straight edge.  The white lines are streamlines, while the dashed black lines represent the sonic surface.}
\label{fig:2D-slice}
\end{figure*}

Our final and most sophisticated model utilizes the approach outlined in \citet{wang_2018}, which combines ray-tracing radiative transfer, real-time non-equilibrium thermochemistry, and hydrodynamics based on the higher-order Godunov method code \verb|Athena++| \citep{stone_2020}. Photoevaporation is inherently not a spherically symmetric phenomenon, since it is driven by stellar XUV flux and the star is only in one direction.  Compared to 1D LTE models, this axisymmetric (2.5D) model better captures the anisotropy of the outflow pattern, while the non-LTE thermochemistry self-consistently predicts the mass loss rate and the line profiles. Like \citet{oklopcic_2018} and \cite{wang_2020}, metastable helium is added as a chemical species and key reactions that form and destroy this species are included in the thermochemical network.  The model incorporates a total of 26 species and 135 reactions, including various relevant heating and cooling processes (e.g. photoionization, photodissociation of atomic hydrogen and Lyman-$\alpha$ cooling).

Our simulations are done in a spherical coordinate system centered on the planet, with the polar axis pointing from the center of the planet to the host star. The simulation domain are in the radial and polar directions [r,$\theta$], while symmetry is assumed in the $\phi$ direction. Photons for the ray-tracing calculation are divided into five energy bins: (1) $h\nu = 2~\eV$ for infrared, optical and near ultraviolet (NUV) photons, (2) $h\nu = 7~\eV$ for ``soft'' far ultraviolet (FUV) photons, (3) $h\nu = 12~\eV$ for the Lyman-Werner band FUV photons that can photodissociate molecular hydrogen but cannot ionize them, (4) $h\nu = 20~\eV$ for ``soft'' extreme ultraviolet (soft EUV) photons that can ionize hydrogen but {\it not} helium, (5) $h\nu = 40~\eV$ for hard EUV (and soft X-ray; denoted by ``hard EUV'' hereafter for simplicity) photons that ionize hydrogen {\it and} helium. Photon fluxes in each energy bin are determined according to the planet's orbital separation and the corresponding luminosities for a typical K star, according to the reviews in \cite{oklopcic_2019}.  The EUV flux for this model star is 10,926 erg s$^{-1}$ \AA$^{-1}$, 30\% higher than what we adopted in the 1D TPCI simulations--a negligible difference given the inherent uncertainty in EUV flux.  In addition to the opacities caused by photochemical reactions, we also added an effective opacity
term to all bands, particularly in the optical band $h\nu=2~\eV$ where our opacity calculation did not include the Thomson cross-section $\sigma/\chem{H}\simeq 6.7\times 10^{-25}~\cm^2$.

\begin{deluxetable}{lr}
  \tablecolumns{2} 
  \tabletypesize{\scriptsize}
  \tablewidth{0pt}
  \tablecaption
  {Setups of the fiducial 2.5D numerical models of the evaporating
    atmosphere \label{table:simulation-fiducial} }
  \tablehead{
    \colhead{Item} &
    \colhead{Value}
  }
  \startdata
  Simulation domain & \\
  Radial range & $1.89 \le (r/R_\oplus) \le \ 20$\\
  Latitudinal range & $0\le\theta\le\pi$ \\
  Resolution $(N_{\log r}\times N_{\theta} )$ &
  $144\times 96$ \\ 
  \\
  Atmospheric properties$^*$ & \\
  $\rho(R_\p)$ & $10^{-7}~\g~\cm^{-3}$ \\
  $T(R_\p)\simeq T_{\rm eq}$ & $1990~\K$ \\
  \\
  Radiation flux [photon~$\cm~^{-2}~\s^{-1}$] & \\
  $2~\eV$ (IR/optical)  & $1.1\times 10^{21}$ \\
  $7~\eV$ (Soft FUV)   & $4.3\times 10^{16}$ \\
  $12~\eV$ (Lyman-Werner FUV) & $1.5\times 10^{13}$ \\
  $20~\eV$ (Soft EUV)  & $1.2\times 10^{13}$ \\
  $40~\eV$ (Hard EUV)  & $1.6\times 10^{14}$ \\
  \\
  Initial abundances [$n_{\chem{X}}/n_{\rm atom}$] $^\dagger$& \\
  \chem{H_2} & 0.455\\
  He & 0.091\\
  \chem{H_2O} & $1.8 \times 10^{-4}$\\
  CO & $1.4 \times 10^{-4}$\\
  S  & $2.8 \times 10^{-5}$\\
  Si & $1.7 \times 10^{-6}$\\
  \enddata
  \tablecomments{$*$: $R_\p\simeq 1.89~R_\oplus$ is the
    size of the rocky planet core.
    \\
    $\dagger$: Dust grains are not included,
    since $T_{\rm eq}\simeq 1990~\K$ is well above the dust
    sublimation temperature. }
\end{deluxetable}

We note that the models we present here do not account for the effect of the stellar wind on the predicted mass loss rate.  In previously published 1D models (i.e. \citealt{murray-clay_2009}), the stellar wind acts as a simple scalar suppression force. In \cite{wang_2020}, they discuss the effect of the stellar wind on planetary outflows using the same hydrodynamic model presented here. They find that in higher dimensions, the stellar wind only suppresses the day side mass loss; the outflow is simply redirected towards the night side of the planet, forming a comet-like tail. In these models, adding a solar-like stellar wind (we note that 55 Cnc e is an old, sun-like star) only changes the 3D mass loss rate by a few percent, corresponding to a change in the equivalent width of He absorption of less than 10\% \cite[see Table 2 in][]{wang_2020}. In short, stellar wind may alter the line profile of He absorption, but it is unable to quench the outflow in three dimensions and is therefore unlikely to be the reason behind the non-detection of He absorption for 55 Cnc e.

A typical planetary atmosphere consists of a convective
interior and a quasi-isothermal exterior
\citep[e.g.][]{rafikov_2006}, but our numerical tests
found that even a thin convective isentropic layer will
cause the whole atmosphere to become overdense and unbounded for such
a close-in low-mass planet (the boil-off regime, see discussions in
\citealt{owen_2016} and \citealt{wang_2018}). We therefore set up
the model atmospheres with a quasi-isothermal layer directly above the rocky core. In contrast, the TPCI model did not set a solid surface or impose a truncation of the gas reservoir; instead, it sets pressure and density boundary conditions at the inner radius. We summarize the key quantities that define the 2.5D fiducial model in Table~\ref{table:simulation-fiducial}. In
addition to the fiducial model, which has a density of $\rho(R_\p) =
10^{-7}~\g~\cm^{-3}$ at the rocky surface, we also considered
models with $\rho(R_\p) =
10^{-5}~\g~\cm^{-3}$ and $10^{-9}~\g~\cm^{-3}$,
respectively. As with the TPCI models, we also considered a scenarios with a He-dominated atmosphere (1\% hydrogen and 99\% helium by atom number) spanning the same three surface pressures.

Figure \ref{fig:2D-slice} shows the 2D profiles of density, temperature, and triplet helium density, along with streamlines and the sonic surface.  As can be seen, the anti-stellar side of the planet has drastically different physical conditions from the star-facing side.  However, due to its compactness, much of colder, less dense material on the anti-stellar side would not block any more light during transit than the planet itself.  Except for this region, the rest of the simulation domain is largely spherically symmetric, especially in the number density of triplet helium, which directly determines the helium absorption signature.

\begin{figure}
  \centering
  \includegraphics[width=3.4in, keepaspectratio]
  {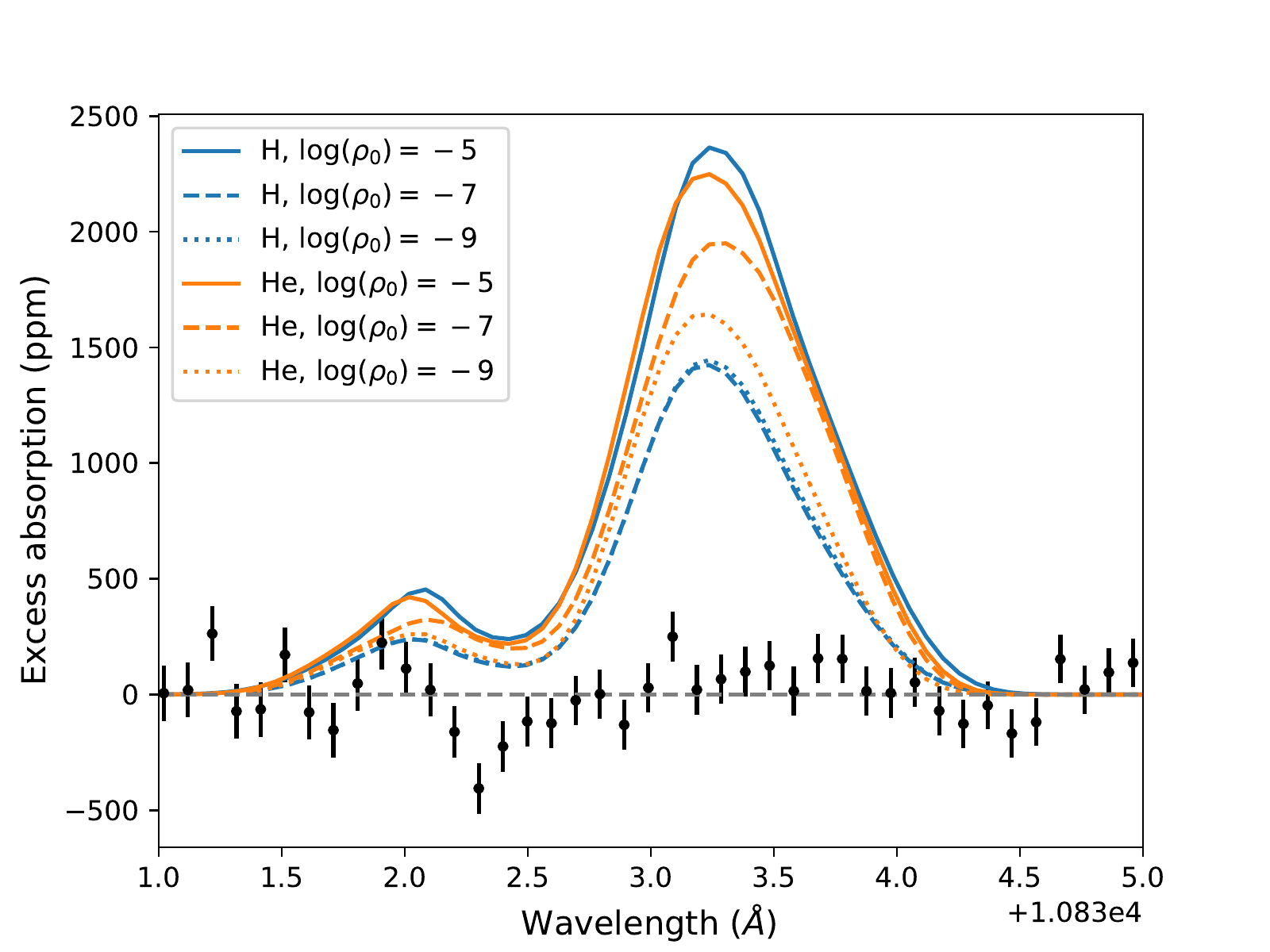}
  \caption{Metastable helium transmission spectra of model
    atmospheres for the simulations described in
    \S\ref{sec:simulations} and Tables~
    \ref{table:simulation-fiducial},
    \ref{table:simulation-results}, presenting the
    time-averaged (from the end of ingress through the start
    of egress) excess absorption. The horizontal dashed line
    indicates zero excess absorption for
    reference. Different reference densities (measured at
    the planet surface $R_{\rm p}$) are marked by different
    line styles, with $log(\rho_0)=-5, -7, -9$ corresponding to roughly 500, 5, and 0.05 millibar respectively. Orange lines represent helium-dominated models, while blue lines represent hydrogen-dominated models.}
  \label{fig:simulation-spec}
\end{figure}

Figure~\ref{fig:tpci_models_comp} shows the radial
profiles of temperature, velocity, hydrogen number density, triplet helium number density, neutral H fraction, and neutral He fraction for the fiducial model (hydrogen dominated, $\rho(R_p)=10^{-7}$ g cm$^{-3}$), showing a fairly
typical supersonic photoevaporative outflow that carries
metastable helium atoms. The assumed EUV flux of 55 Cnc e produces a relative abundance of metastable helium in 55 Cnc e's atmosphere of $\sim 10^{-7}$, nearly identical to the abundance predicted by TPCI. Despite
this low abundance, the transmission spectra in
Figure~\ref{fig:simulation-spec} still have clearly recognizable
 excess absorption with amplitudes of a few thousand ppm. The absorption is
still greater than 1000 ppm even for the thinnest atmosphere ($\rho(R_\p) = 10^{-9}~\g~\cm^{-3}$). The mass-loss timescales of these models, summarized in
Table~\ref{table:simulation-results}, range between years and thousands of years. The mass loss rates are within 50\% of those predicted by TPCI for both hydrogen and helium dominated atmospheres. These mass-loss rates are time-averaged after the simulation reaches a quasi-steady state after dozens of dynamical timescales elapsed since the start of the simulations. The non-detection of helium absorption is therefore in good agreement with the short dispersal timescales for these atmospheres, implying that 55 Cnc e would not have been able to retain a primordial atmosphere for very long after the dispersal of the gas disk.

\begin{deluxetable*}{lccccccc}
  \tablecolumns{7}
  \tabletypesize{\scriptsize}
  \tablewidth{1.0\columnwidth} 
  \tablecaption{2.5D Simulations: Parameters and Results
    \label{table:simulation-results}
  }
 \tablehead{
    \colhead{Model} \vspace{-0.25cm} &
    \colhead{$\rho(R_{\rm p})$} &
    \colhead{$M_{env}/M_p$} &
    \colhead{$P(R_p)$} &
    \colhead{$\dot{M}$} &
    \colhead{$\tau_{\rm disp}^a$} &
    \colhead{$\mean{W_\lambda}^b$} &
    \colhead{$\Delta \ln(L)^c$}
    \\
    \colhead{No.} &
    \colhead{($\g~\cm^{-3}$)} &
    \colhead{} &
    \colhead{(mbar)} &
    \colhead{($10^{10}~g~s^{-1}$)} &    
    \colhead{($\yr$)} &
    \colhead{($10^{-3}~\ang$)} &
    \colhead{} 
  }
   \startdata
   H/He Envelope\\
   1 & $10^{-5}$ & $0.98\times 10^{-7}$ & 500 & $1.5$ & $5000$ & $2.4$ & $748$\\
   2$^*$ & $10^{-7}$ & $0.98\times 10^{-9}$ & 5 & $0.84$ & $86$ & $1.4$ & $243$\\
   3 & $10^{-9}$ & $0.98\times 10^{-11}$ & 0.05 & $0.85$ & $0.9$ & $1.4$ & $252$\\
   \hline
   He-Dominated\\
   4 & $10^{-5}$ & $0.80\times 10^{-7}$ & 400 & $1.1$ & $4000$ & $3.3$ & $712$\\
   5 & $10^{-7}$ & $0.80\times 10^{-9}$ & 4 & $0.94$ & $65$ & $2.0$ & $538$\\
   6 & $10^{-9}$ & $0.80\times 10^{-11}$ & 0.04 & $0.70$ & $0.9$ & $1.6$ & $342$\\    
   \enddata
   \tablecomments{$a$: Dispersal timescale, defined by
   the current atmospheric mass divided by the steady-state
   mass loss rate.
   \\
   $b$: Dimensional equivalent width of excess
   absorption, time-averaged from the end of ingress through
   the start of egress.
   \\
   $c$: Difference in log likelihood between this model and a model with zero absorption.  A bigger number means a worse fit.
   \\
   $*$: Fiducial model.
   (\S\ref{sec:simulations}).}
\end{deluxetable*}

\section{Discussion}
\label{sec:discussion}
Now that we have explored the limits that can be placed on 55 Cnc e's atmosphere from our helium observations alone, we next consider how these results relate to other published observations of 55 Cnc e's atmosphere.  We first focus on \cite{ehrenreich_2012}, which presents a Lyman $\alpha$ transit of 55 Cnc e from HST/STIS.  Lyman $\alpha$ and the helium 1083 nm line are both good probes of the outflowing atmosphere, but the former is sensitive only to the high speed portions of the outflow while the latter can only probe the low-speed portions of the outflow where metastable helium exists.  \cite{ehrenreich_2012} measures the transit depth between 1215.36 \ang  and 1215.67 \ang, obtaining a value of $0.3 \pm 2.4$\%.  From this, they constrain the mass loss rate to be below 3\e{8} g/s (3$\sigma$).  They interpret their data using the \cite{bourrier_2013} model, which is a 3D particle-based simulation that takes into account radiation pressure, photoionization, the stellar wind, and the effects of self-shielding for both stellar photons and protons.  The inclusion of outward forces from radiation pressure and stellar wind gives the outflow a trailing cometary tail: a consequence of the Coriolis acceleration $-2\mathbf{\Omega} \times \mathbf{v}$.  This model has 5 tunable parameters: the mass loss rate, the stellar EUV flux, and three stellar wind parameters: bulk velocity, temperature, and density.  (By contrast, our 2.5D model is a hydrodynamics code which predicts the mass loss rate given the EUV flux, but does not take the stellar wind into account and cannot model a tail.)  As with all Lyman alpha observations of even the nearest stars, the line core is completely absorbed by the interstellar medium, and only the far wings are visible.  This means that the detectability of an absorption signal depends as much on the highly uncertain kinematic structure of the outflow as on the quantity of outflowing gas.

Even though we quote a slightly higher upper limit on the mass loss rate ($\sim10^9$ g/s from the isothermal Parker wind model) than \cite{ehrenreich_2012} (3\e{8} g/s), our models are in fact consistent with their Lyman alpha non-detection even at mass loss rates that are definitively ruled out by our helium non-detection. Figure 12 of \cite{salz_2016} compares the Lyman alpha data to their TPCI simulation of 55 Cnc e under the assumption of a hydrogen-dominated atmosphere, finding that the two are consistent.  As we have seen, however, the helium absorption signal predicted by TPCI for the hydrogen-dominated atmosphere case is highly inconsistent with our non-detection.  In addition, a 1D model is not ideal for modelling Lyman-$\alpha$ because the strength of the line absorption, combined with interstellar absorption that eliminates the line core, make the kinematic structure of the outflow crucially important in predicting the observable signal.  For this reason, we used the same code we used to run our 2.5D models to run a helium-dominated 3D hydrodynamic model (resembling model 5 in Table \ref{table:simulation-results}, corresponding to an intermediate mass helium atmosphere).  We found that just 1\% hydrogen in the outflow is enough to give a 12\% excess absorption depth in the Lyman $\alpha$ line core, but the line core is not observable due to interstellar absorption.  

\begin{figure}[ht]
  \includegraphics
    [width=0.5\textwidth]{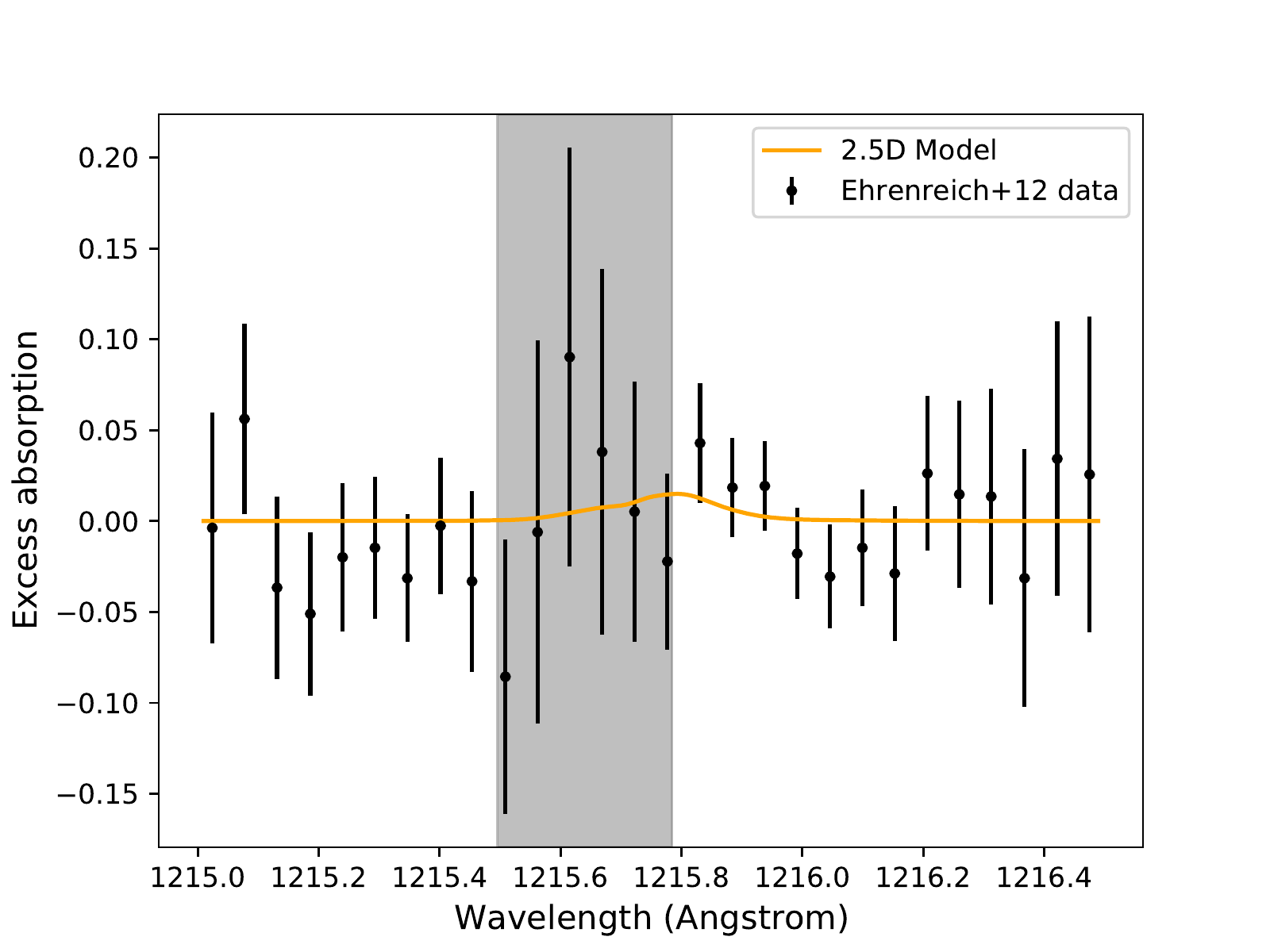}
    \caption{Lyman alpha observations of \cite{ehrenreich_2012}, compared to predictions of our helium-dominated 2.5D model (model 5 in Table \ref{table:simulation-results}).  Within the gray shaded region, the intrinsic stellar flux is fully absorbed by the ISM, but the observed flux is non-zero because of the broad instrumental line spread profile.}
\label{fig:ly_alpha_model_comp}
\end{figure}

To determine what is observable, we start with the stellar intrinsic Lyman alpha profile, corrected for interstellar absorption, that \cite{ehrenreich_2012} provide in Figure 6.  We model the observed out-of-transit spectrum by convolving the absorbed profile with the line spread function of STIS, as provided by STScI\footnote{\url{https://www.stsci.edu/hst/instrumentation/stis/performance/spectral-resolution}}.  We model the observed in-transit spectrum by multiplying the post-ISM-absorption profile by the predicted Lyman alpha absorption profile from the exosphere and convolving the product with the line spread profile.  The resulting excess absorption spectrum for our helium-dominated 2.5D model is shown in Figure \ref{fig:ly_alpha_model_comp}. We did not show the hydrogen-dominated case as the high-opacity region extends well beyond the domain of our simulation. It is clear that the predicted Lyman alpha absorption is fully consistent with the observations.  As a caveat, we note that absorption is often seen in the high-velocity wings of Lyman-$\alpha$ observations, which could be due to charge exchange with the stellar wind \citep{holmstrohm_2008,tremblin_2013}.  Since our model does not include stellar wind, it may severely underestimate the absorption at observable wavelengths.  Were the stellar wind to be included, it is possible that the Lyman alpha non-detection would become a stronger constraint on mass loss than the helium non-detection.

As noted in the previous section, the observations by \cite{ehrenreich_2012} coincided with the stellar activity minimum.  As far as we know, nobody observed 55 Cnc e in either Lyman alpha or He I 1083 nm during the stellar maximum in mid 2017.  Our observations took place in between said maximum and the next minimum in late 2022.  The star's X-ray luminosity is 2.4 times higher in this phase than it was in 2012, making it conceivable that any potential outflow from the planet would have increased in strength relative to the epoch of the Lyman $\alpha$ observations during the stellar minimum.  Both the TPCI models and the 2.5D simulations assume the higher X-ray flux relevant for the epoch of our helium observations, and as a result they may over-predict the magnitude of the Lyman-$\alpha$ absorption at the epoch of the HST observations.  Since the predicted Lyman-$\alpha$ absorption from these higher X-ray flux models is already undetectable, there is no need to recalculate the models with a lower X-ray flux level.

Prior to our observations, it was also conceivable that 55 Cnc e might have had a helium-dominated atmosphere.  However, our observations show no evidence of helium absorption, and our 2.5D simulations show that the mass loss rates are much too fast for the planet to keep either a hydrogen- or helium-dominated atmosphere.  Fast mass loss also makes it difficult to create a helium-dominated atmosphere in the first place, because escape rates much faster than the diffusion-limited rate do not significantly fractionate the elements \citep{hu_2015}.

Our observations appear to contradict those of \cite{tsiaras_2016}, who analyzed two transits of 55 Cnc e observed with HST/WFC3, and found an upward-sloping spectral feature consistent with HCN absorption in a lightweight atmosphere.  They perform a free retrieval on the transmission spectrum with TauREx and find a mean molecular weight $\mu$ of 2--6 (their Figure 9).  The atmospheric scale height is inversely proportional to $\mu$; thus, the amplitude of features in a transmission spectrum is also inversely proportional to $\mu$.  $\mu=2-6$ is consistent with a hydrogen dominated atmosphere ($\mu=2.3$) or a helium dominated atmosphere ($\mu \approx 4$), but excludes atmospheres dominated by heavier molecules (i.e. water with $\mu=18$, N$_2$ with $\mu=14$, O$_2$ with $\mu=16$).  If the atmosphere is not hydrogen or helium dominated, absorption features from any molecule, including HCN, would be undetectable.  If $\mu=16$, for example, the scale height would be 47 km (assuming equilibrium temperature) and the change in transit depth corresponding to one scale height would be 3 ppm.  This lies far below the detection threshold of the observations reported in \cite{tsiaras_2016}.  We conclude that our non-detection of escaping helium, together with the non-detection of escaping hydrogen by \cite{ehrenreich_2012} and the strong irradiation of the planet over its long life, make it unlikely that the features detected by Tsiaras et al. in the HST/WFC3 spectrum are planetary in nature.

Two more pieces of evidence complicate the picture: the significant phase offset in the Spitzer 4.5 $\mu m$ phase curve \citep{demory_2016b}, and the year-to-year variability in the secondary eclipse depth in this same band \citep{demory_2016a}.  To explain the eclipse variability, one could invoke a magma world with no atmosphere, or an extremely tenuous mineral atmosphere in equilibrium with the molten surface \citep{ito_2015}, in which case ejecta from volcanic eruptions could periodically shroud the surface.  The mass and radius of 55 Cnc e are consistent with a world without an atmosphere \citep{bourrier_2018}.  However, the phase offset suggests a thick, high molecular weight atmosphere.  This point was discussed in \cite{angelo_2017}, who suggested that an N$_2$-dominated atmosphere would be consistent with the Spitzer data.  The possible composition of a nitrogen-dominated atmosphere and the observability of spectral features are explored in \cite{miguel_2019} and \cite{zilinskas_2020}.  However, such a thick atmosphere would be unlikely to change substantially on year-long timescales.  Our non-detection of helium is agnostic to all potential high mean molecular weight atmospheres, as helium is not expected to be a significant component of secondary (e.g., non-primordial) atmospheres.  While photodissociation at the top of a water-rich atmosphere might create a detectable Lyman-$\alpha$ signal, high resolution spectral observations rule out water-rich atmospheres (volume mixing ratio $>$ 0.1\%) with a mean molecular weight of $<=$ 15 g/mol at a 3$\sigma$ confidence level \citep{jindal_2020}.

\section{Conclusion}
We observed two transits of 55 Cnc e using Keck/NIRSPEC to look for metastable helium absorption in the 1083 nm line.  We found no absorption greater than 250 ppm (90\% upper limit), and used three independent models to interpret this result.  First, an isothermal Parker wind model \citep{oklopcic_2018} puts constraints on the temperature and mass loss rate (Figure \ref{fig:mass_loss_constraint}), with the mass loss rate constrained to less than $\sim10^9$ g/s for exosphere temperatures of 5000-6000 K.  This exosphere temperature is obtained from our second model, The PLUTO-\cloudy Interface (TPCI), which can model the outflow in a 1D fashion given the stellar XUV spectrum.  TPCI predicts a mass loss rate of 1.1\e{10} g/s for a hydrogen-dominated atmosphere and 7.5\e{9} g/s for a helium-dominated atmosphere, both of which result in absorption several times stronger than what is (not) observed (Figure \ref{fig:tpci_obs_comparison}).  Our third and most sophisticated model is a 2.5D model \citep{wang_2018} which combines ray-tracing radiative transfer, real-time non-equilibrium thermochemistry, and hydrodynamics to model the outflow, assuming it is symmetric about the star-planet axis.  Even for extremely thin atmospheres with dispersal timescales of millenia or less, the model still predicts high mass loss rates of $\sim10^{10}$ g/s for both hydrogen and helium dominated atmospheres (see Table \ref{table:simulation-results}), which result in 1500--2500 ppm excess absorption--many times higher than what is observed (Figure \ref{fig:simulation-spec}).  Although the significant model uncertainties must be kept in mind, our observations provide strong evidence against the existence of a low mean molecular weight primordial atmosphere on 55 Cnc e.

If 55 Cnc e instead possesses a high mean molecular weight secondary atmosphere, detection via transit spectroscopy will be extremely challenging.  Ultimately, we believe that emission spectroscopy with next-generation telescopes--JWST, TMT, ELT, and GMT--represents the best path forward.  This planet's high dayside temperature ($\sim$2700 K) makes it a particularly favorable target for emission spectroscopy, and the magnitude of spectral features seen in emission is independent of the mean molecular weight of the atmosphere.  While the brightness of 55 Cnc makes it feasible to search for relatively small signals, systematics that are negligible for low SNR targets become important when the photon noise is small.  In this paper, we saw that crosstalk, fringing, and the inseparability of the 2D PSF each required special handling; observations with both JWST and next-generation ground-based telescopes will likely encounter similar technical challenges.  Despite these challenges, atmosphere modeling studies suggest that it is indeed possible to detect high mean molecular weight atmospheres for 55 Cnc e using next-generation telescopes (e.g., \citealt{zilinskas_2020}).  To date, atmospheric absorption features have only been detected for planets with relatively massive, hydrogen-rich atmospheres.  Detecting a high mean molecular weight atmosphere around a high density planet like 55 Cnc e would provide invaluable insights into the nature and origin of the broader population of short-period super-Earths.

\textit{Software:}  \texttt{numpy \citep{van_der_walt_2011}, scipy \citep{virtanen_2020}, matplotlib \citep{hunter_2007}, dynesty \citep{speagle_2019}, corner \citep{foreman-mackey_2016}}

\section{Acknowledgments}
M.Z. would like to acknowledge Joe Ninan, who was indispensable in deciphering the mysteries of NIRSPEC systematics.  He also acknowledges Jacob Bean for sending most of the data plotted in Figure \ref{fig:context}.

\bibliographystyle{apj} \bibliography{main}

\begin{thebibliography}{72}
\expandafter\ifx\csname natexlab\endcsname\relax\def\natexlab#1{#1}\fi

\bibitem[{{Angelo} \& {Hu}(2017)}]{angelo_2017}
{Angelo}, I., \& {Hu}, R. 2017, \aj, 154, 232

\bibitem[{{Bolton} \& {Schlegel}(2010)}]{bolton_2010}
{Bolton}, A.~S., \& {Schlegel}, D.~J. 2010, \pasp, 122, 248

\bibitem[{{Bourrier} \& {Lecavelier des Etangs}(2013)}]{bourrier_2013}
{Bourrier}, V., \& {Lecavelier des Etangs}, A. 2013, \aap, 557, A124

\bibitem[{{Bourrier} {et~al.}(2018){Bourrier}, {Dumusque}, {Dorn}, {Henry},
  {Astudillo-Defru}, {Rey}, {Benneke}, {H{\'e}brard}, {Lovis}, {Demory},
  {Moutou}, \& {Ehrenreich}}]{bourrier_2018}
{Bourrier}, V., {Dumusque}, X., {Dorn}, C., {et~al.} 2018, \aap, 619, A1

\bibitem[{{Chachan} \& {Stevenson}(2018)}]{chachan_2018}
{Chachan}, Y., \& {Stevenson}, D.~J. 2018, \apj, 854, 21

\bibitem[{{Dai} {et~al.}(2019){Dai}, {Masuda}, {Winn}, \& {Zeng}}]{Dai}
{Dai}, F., {Masuda}, K., {Winn}, J.~N., \& {Zeng}, L. 2019, \apj, 883, 79

\bibitem[{{Demory} {et~al.}(2016{\natexlab{a}}){Demory}, {Gillon},
  {Madhusudhan}, \& {Queloz}}]{demory_2016a}
{Demory}, B.-O., {Gillon}, M., {Madhusudhan}, N., \& {Queloz}, D.
  2016{\natexlab{a}}, \mnras, 455, 2018

\bibitem[{{Demory} {et~al.}(2016{\natexlab{b}}){Demory}, {Gillon}, {de Wit},
  {Madhusudhan}, {Bolmont}, {Heng}, {Kataria}, {Lewis}, {Hu}, {Krick},
  {Stamenkovi{\'c}}, {Benneke}, {Kane}, \& {Queloz}}]{demory_2016b}
{Demory}, B.-O., {Gillon}, M., {de Wit}, J., {et~al.} 2016{\natexlab{b}}, \nat,
  532, 207

\bibitem[{Drake {et~al.}(2020)Drake, Kashyap, Wargelin, \& Wolk}]{drake_2020}
Drake, J.~J., Kashyap, V.~L., Wargelin, B.~J., \& Wolk, S.~J. 2020, The
  Astrophysical Journal, 893, 137

\bibitem[{{Eastman} {et~al.}(2013){Eastman}, {Gaudi}, \& {Agol}}]{Eastman2013}
{Eastman}, J., {Gaudi}, B.~S., \& {Agol}, E. 2013, \pasp, 125, 83

\bibitem[{{Ehrenreich} {et~al.}(2012){Ehrenreich}, {Bourrier}, {Bonfils},
  {Lecavelier des Etangs}, {H{\'e}brard}, {Sing}, {Wheatley}, {Vidal-Madjar},
  {Delfosse}, {Udry}, {Forveille}, \& {Moutou}}]{ehrenreich_2012}
{Ehrenreich}, D., {Bourrier}, V., {Bonfils}, X., {et~al.} 2012, \aap, 547, A18

\bibitem[{{Ferland} {et~al.}(2013){Ferland}, {Porter}, {van Hoof}, {Williams},
  {Abel}, {Lykins}, {Shaw}, {Henney}, \& {Stancil}}]{ferland_2013}
{Ferland}, G.~J., {Porter}, R.~L., {van Hoof}, P.~A.~M., {et~al.} 2013, \rmxaa,
  49, 137

\bibitem[{{Foreman-Mackey}(2016)}]{foreman-mackey_2016}
{Foreman-Mackey}, D. 2016, The Journal of Open Source Software, 1, 24

\bibitem[{{Fulton} {et~al.}(2017){Fulton}, {Petigura}, {Howard}, {Isaacson},
  {Marcy}, {Cargile}, {Hebb}, {Weiss}, {Johnson}, {Morton}, {Sinukoff},
  {Crossfield}, \& {Hirsch}}]{fulton_2017}
{Fulton}, B.~J., {Petigura}, E.~A., {Howard}, A.~W., {et~al.} 2017, \aj, 154,
  109

\bibitem[{{Gaidos} {et~al.}(2020){Gaidos}, {Hirano}, {Wilson}, {France},
  {Rockcliffe}, {Newton}, {Feiden}, {Krishnamurthy}, {Harakawa}, {Hodapp},
  {Ishizuka}, {Jacobson}, {Konishi}, {Kotani}, {Kudo}, {Kurokawa}, {Kuzuhara},
  {Nishikawa}, {Omiya}, {Serizawa}, {Tamura}, {Ueda}, \&
  {Vievard}}]{gaidos_2020b}
{Gaidos}, E., {Hirano}, T., {Wilson}, D.~J., {et~al.} 2020, \mnras, 498, L119

\bibitem[{Gaidos {et~al.}(2020)Gaidos, Hirano, Mann, Owens, Berger, France,
  Vanderburg, Harakawa, Hodapp, Ishizuka, Jacobson, Konishi, Kotani, Kudo,
  Kurokawa, Kuzuhara, Nishikawa, Omiya, Serizawa, Tamura, \&
  Ueda}]{gaidos_2020a}
Gaidos, E., Hirano, T., Mann, A.~W., {et~al.} 2020, Monthly Notices of the
  Royal Astronomical Society, 495, 650

\bibitem[{{George} {et~al.}(2020){George}, {Tulloch}, {Ives}, \& {ESO Detector
  Group}}]{george_2020}
{George}, E.~M., {Tulloch}, S.~M., {Ives}, D.~J., \& {ESO Detector Group}.
  2020, Journal of Astronomical Telescopes, Instruments, and Systems, 6, 011003

\bibitem[{{Ginzburg} {et~al.}(2018){Ginzburg}, {Schlichting}, \&
  {Sari}}]{ginzburg_2018}
{Ginzburg}, S., {Schlichting}, H.~E., \& {Sari}, R. 2018, \mnras, 476, 759

\bibitem[{{Hirano} {et~al.}(2020){Hirano}, {Krishnamurthy}, {Gaidos},
  {Flewelling}, {Mann}, {Narita}, {Plavchan}, {Kotani}, {Tamura}, {Harakawa},
  {Hodapp}, {Ishizuka}, {Jacobson}, {Konishi}, {Kudo}, {Kurokawa}, {Kuzuhara},
  {Nishikawa}, {Omiya}, {Serizawa}, {Ueda}, \& {Vievard}}]{hirano_2020}
{Hirano}, T., {Krishnamurthy}, V., {Gaidos}, E., {et~al.} 2020, \apjl, 899, L13

\bibitem[{{Holmstr{\"o}m} {et~al.}(2008){Holmstr{\"o}m}, {Ekenb{\"a}ck},
  {Selsis}, {Penz}, {Lammer}, \& {Wurz}}]{holmstrohm_2008}
{Holmstr{\"o}m}, M., {Ekenb{\"a}ck}, A., {Selsis}, F., {et~al.} 2008, \nat,
  451, 970

\bibitem[{{Horne}(1986)}]{horne_1986}
{Horne}, K. 1986, \pasp, 98, 609

\bibitem[{{Hu} {et~al.}(2015){Hu}, {Seager}, \& {Yung}}]{hu_2015}
{Hu}, R., {Seager}, S., \& {Yung}, Y.~L. 2015, \apj, 807, 8

\bibitem[{{Hunter}(2007)}]{hunter_2007}
{Hunter}, J.~D. 2007, Computing in Science and Engineering, 9, 90

\bibitem[{{Husser} {et~al.}(2013){Husser}, {Wende-von Berg}, {Dreizler},
  {Homeier}, {Reiners}, {Barman}, \& {Hauschildt}}]{husser_2013}
{Husser}, T.-O., {Wende-von Berg}, S., {Dreizler}, S., {et~al.} 2013, \aap,
  553, A6

\bibitem[{{Ionov} \& {Shematovich}(2015)}]{ionov_2015}
{Ionov}, D.~E., \& {Shematovich}, V.~I. 2015, Solar System Research, 49, 339

\bibitem[{{Ito} {et~al.}(2015){Ito}, {Ikoma}, {Kawahara}, {Nagahara},
  {Kawashima}, \& {Nakamoto}}]{ito_2015}
{Ito}, Y., {Ikoma}, M., {Kawahara}, H., {et~al.} 2015, \apj, 801, 144

\bibitem[{{Jindal} {et~al.}(2020){Jindal}, {de Mooij}, {Jayawardhana},
  {Deibert}, {Brogi}, {Rustamkulov}, {Fortney}, {Hood}, \&
  {Morley}}]{jindal_2020}
{Jindal}, A., {de Mooij}, E. J.~W., {Jayawardhana}, R., {et~al.} 2020, \aj,
  160, 101

\bibitem[{{Kasper} {et~al.}(2020){Kasper}, {Bean}, {Oklop{\v{c}}i{\'c}},
  {Malsky}, {Kempton}, {D{\'e}sert}, {Rogers}, \& {Mansfield}}]{kasper_2020}
{Kasper}, D., {Bean}, J.~L., {Oklop{\v{c}}i{\'c}}, A., {et~al.} 2020, \aj, 160,
  258

\bibitem[{{Kim} {et~al.}(2015){Kim}, {Prato}, \& {McLean}}]{kim_2015}
{Kim}, S., {Prato}, L., \& {McLean}, I. 2015, {REDSPEC: NIRSPEC data reduction}

\bibitem[{{Kreidberg}(2015)}]{Kreidberg}
{Kreidberg}, L. 2015, \pasp, 127, 1161

\bibitem[{{Lee} \& {Chiang}(2017)}]{lee_2017}
{Lee}, E.~J., \& {Chiang}, E. 2017, \apj, 842, 40

\bibitem[{{Linsky} {et~al.}(2014){Linsky}, {Fontenla}, \&
  {France}}]{linsky_2014}
{Linsky}, J.~L., {Fontenla}, J., \& {France}, K. 2014, \apj, 780, 61

\bibitem[{{Lopez} \& {Fortney}(2013)}]{lopez_2013}
{Lopez}, E.~D., \& {Fortney}, J.~J. 2013, \apj, 776, 2

\bibitem[{{Lord}(1992)}]{lord_1992}
{Lord}, S.~D. 1992, {A new software tool for computing Earth's atmospheric
  transmission of near- and far-infrared radiation}, NASA Technical Memorandum
  103957

\bibitem[{{Malsky} \& {Rogers}(2020)}]{malsky_2020}
{Malsky}, I., \& {Rogers}, L.~A. 2020, \apj, 896, 48

\bibitem[{{Marsh}(1989)}]{marsh_1989}
{Marsh}, T.~R. 1989, \pasp, 101, 1032

\bibitem[{{Martin} {et~al.}(2018){Martin}, {Fitzgerald}, {McLean}, {Doppmann},
  {Kassis}, {Aliado}, {Canfield}, {Johnson}, {Kress}, {Lanclos}, {Magnone},
  {Sohn}, {Wang}, \& {Weiss}}]{martin_2018}
{Martin}, E.~C., {Fitzgerald}, M.~P., {McLean}, I.~S., {et~al.} 2018, in
  Society of Photo-Optical Instrumentation Engineers (SPIE) Conference Series,
  Vol. 10702, \procspie, 107020A

\bibitem[{{Mazeh} {et~al.}(2007){Mazeh}, {Tamuz}, \& {Zucker}}]{mazeh_2007}
{Mazeh}, T., {Tamuz}, O., \& {Zucker}, S. 2007, Astronomical Society of the
  Pacific Conference Series, Vol. 366, {The Sys-Rem Detrending Algorithm:
  Implementation and Testing}, ed. C.~{Afonso}, D.~{Weldrake}, \& T.~{Henning},
  119

\bibitem[{Mignone {et~al.}(2007)Mignone, Bodo, Massaglia, Matsakos, Tesileanu,
  Zanni, \& Ferrari}]{mignone_2007}
Mignone, A., Bodo, G., Massaglia, S., {et~al.} 2007, The Astrophysical Journal
  Supplement Series, 170, 228

\bibitem[{{Miguel}(2019)}]{miguel_2019}
{Miguel}, Y. 2019, \mnras, 482, 2893

\bibitem[{{Murray-Clay} {et~al.}(2009){Murray-Clay}, {Chiang}, \&
  {Murray}}]{murray-clay_2009}
{Murray-Clay}, R.~A., {Chiang}, E.~I., \& {Murray}, N. 2009, \apj, 693, 23

\bibitem[{{Ninan} {et~al.}(2020){Ninan}, {Stefansson}, {Mahadevan}, {Bender},
  {Robertson}, {Ramsey}, {Terrien}, {Wright}, {Diddams}, {Kanodia}, {Cochran},
  {Endl}, {Ford}, {Fredrick}, {Halverson}, {Hearty}, {Jennings}, {Kaplan},
  {Lubar}, {Metcalf}, {Monson}, {Nitroy}, {Roy}, \& {Schwab}}]{ninan_2020}
{Ninan}, J.~P., {Stefansson}, G., {Mahadevan}, S., {et~al.} 2020, \apj, 894, 97

\bibitem[{{Nortmann} {et~al.}(2018){Nortmann}, {Pall{\'e}}, {Salz},
  {Sanz-Forcada}, {Nagel}, {Alonso-Floriano}, {Czesla}, {Yan}, {Chen},
  {Snellen}, {Zechmeister}, {Schmitt}, {L{\'o}pez-Puertas}, {Casasayas-Barris},
  {Bauer}, {Amado}, {Caballero}, {Dreizler}, {Henning}, {Lamp{\'o}n}, {Montes},
  {Molaverdikhani}, {Quirrenbach}, {Reiners}, {Ribas}, {S{\'a}nchez-L{\'o}pez},
  {Schneider}, \& {Zapatero Osorio}}]{nortmann_2018}
{Nortmann}, L., {Pall{\'e}}, E., {Salz}, M., {et~al.} 2018, Science, 362, 1388

\bibitem[{{Oklop{\v{c}}i{\'c}}(2019)}]{oklopcic_2019}
{Oklop{\v{c}}i{\'c}}, A. 2019, arXiv e-prints, arXiv:1903.02576

\bibitem[{{Oklop{\v{c}}i{\'c}} \& {Hirata}(2018)}]{oklopcic_2018}
{Oklop{\v{c}}i{\'c}}, A., \& {Hirata}, C.~M. 2018, \apj, 855, L11

\bibitem[{{Owen}(2019)}]{owen_2019}
{Owen}, J.~E. 2019, Annual Review of Earth and Planetary Sciences, 47, 67

\bibitem[{{Owen} \& {Jackson}(2012)}]{owen_2012}
{Owen}, J.~E., \& {Jackson}, A.~P. 2012, \mnras, 425, 2931

\bibitem[{{Owen} \& {Wu}(2016)}]{owen_2016}
{Owen}, J.~E., \& {Wu}, Y. 2016, \apj, 817, 107

\bibitem[{{Palle, E.} {et~al.}(2020){Palle, E.}, {Nortmann, L.},
  {Casasayas-Barris, N.}, {Lamp\'on, M.}, {L\'opez-Puertas, M.}, {Caballero, J.
  A.}, {Sanz-Forcada, J.}, {Lara, L. M.}, {Nagel, E.}, {Yan, F.},
  {Alonso-Floriano, F. J.}, {Amado, P. J.}, {Chen, G.}, {Cifuentes, C.},
  {Cort\'es-Contreras, M.}, {Czesla, S.}, {Molaverdikhani, K.}, {Montes, D.},
  {Passegger, V. M.}, {Quirrenbach, A.}, {Reiners, A.}, {Ribas, I.},
  {S\'anchez-L\'opez, A.}, {Schweitzer, A.}, {Stangret, M.}, {Zapatero Osorio,
  M. R.}, \& {Zechmeister, M.}}]{palle_2020}
{Palle, E.}, {Nortmann, L.}, {Casasayas-Barris, N.}, {et~al.} 2020, A\&A, 638,
  A61

\bibitem[{{Rafikov}(2006)}]{rafikov_2006}
{Rafikov}, R.~R. 2006, \apj, 648, 666

\bibitem[{{Ridden-Harper} {et~al.}(2016){Ridden-Harper}, {Snellen}, {Keller},
  {de Kok}, {Di Gloria}, {Hoeijmakers}, {Brogi}, {Fridlund}, {Vermeersen}, \&
  {van Westrenen}}]{ridden-harper_2016}
{Ridden-Harper}, A.~R., {Snellen}, I.~A.~G., {Keller}, C.~U., {et~al.} 2016,
  \aap, 593, A129

\bibitem[{{Salz} {et~al.}(2015{\natexlab{a}}){Salz}, {Banerjee}, {Mignone},
  {Schneider}, {Czesla}, \& {Schmitt}}]{salz_2015}
{Salz}, M., {Banerjee}, R., {Mignone}, A., {et~al.} 2015{\natexlab{a}}, \aap,
  576, A21

\bibitem[{{Salz} {et~al.}(2016){Salz}, {Czesla}, {Schneider}, \&
  {Schmitt}}]{salz_2016}
{Salz}, M., {Czesla}, S., {Schneider}, P.~C., \& {Schmitt}, J.~H.~M.~M. 2016,
  \aap, 586, A75

\bibitem[{{Salz} {et~al.}(2015{\natexlab{b}}){Salz}, {Schneider}, {Czesla}, \&
  {Schmitt}}]{salz_2015b}
{Salz}, M., {Schneider}, P.~C., {Czesla}, S., \& {Schmitt}, J.~H.~M.~M.
  2015{\natexlab{b}}, \aap, 576, A42

\bibitem[{{Salz} {et~al.}(2018){Salz}, {Czesla}, {Schneider}, {Nagel},
  {Schmitt}, {Nortmann}, {Alonso-Floriano}, {L{\'o}pez-Puertas}, {Lamp{\'o}n},
  {Bauer}, {Snellen}, {Pall{\'e}}, {Caballero}, {Yan}, {Chen}, {Sanz-Forcada},
  {Amado}, {Quirrenbach}, {Ribas}, {Reiners}, {B{\'e}jar}, {Casasayas-Barris},
  {Cort{\'e}s-Contreras}, {Dreizler}, {Guenther}, {Henning}, {Jeffers},
  {Kaminski}, {K{\"u}rster}, {Lafarga}, {Lara}, {Molaverdikhani}, {Montes},
  {Morales}, {S{\'a}nchez-L{\'o}pez}, {Seifert}, {Zapatero Osorio}, \&
  {Zechmeister}}]{salz_2018}
{Salz}, M., {Czesla}, S., {Schneider}, P.~C., {et~al.} 2018, \aap, 620, A97

\bibitem[{{Sanz-Forcada} {et~al.}(2011){Sanz-Forcada}, {Micela}, {Ribas},
  {Pollock}, {Eiroa}, {Velasco}, {Solano}, \&
  {Garc{\'\i}a-{\'A}lvarez}}]{sanz-forcada_2011}
{Sanz-Forcada}, J., {Micela}, G., {Ribas}, I., {et~al.} 2011, \aap, 532, A6

\bibitem[{{Shaw} \& {Ferland}(2020)}]{shaw_2020}
{Shaw}, G., \& {Ferland}, G.~J. 2020, \mnras, 493, 5153

\bibitem[{{Shematovich} {et~al.}(2014){Shematovich}, {Ionov}, \&
  {Lammer}}]{shematovich_2014}
{Shematovich}, V.~I., {Ionov}, D.~E., \& {Lammer}, H. 2014, \aap, 571, A94

\bibitem[{{Spake} {et~al.}(2018){Spake}, {Sing}, {Evans}, {Oklop{\v{c}}i{\'c}},
  {Bourrier}, {Kreidberg}, {Rackham}, {Irwin}, {Ehrenreich}, {Wyttenbach},
  {Wakeford}, {Zhou}, {Chubb}, {Nikolov}, {Goyal}, {Henry}, {Williamson},
  {Blumenthal}, {Anderson}, {Hellier}, {Charbonneau}, {Udry}, \&
  {Madhusudhan}}]{spake_2018}
{Spake}, J.~J., {Sing}, D.~K., {Evans}, T.~M., {et~al.} 2018, \nat, 557, 68

\bibitem[{Speagle(2020)}]{speagle_2019}
Speagle, J.~S. 2020, Monthly Notices of the Royal Astronomical Society, 493,
  3132

\bibitem[{{Stone} {et~al.}(2020){Stone}, {Tomida}, {White}, \&
  {Felker}}]{stone_2020}
{Stone}, J.~M., {Tomida}, K., {White}, C.~J., \& {Felker}, K.~G. 2020, \apjs,
  249, 4

\bibitem[{{Sulis} {et~al.}(2019){Sulis}, {Dragomir}, {Lendl}, {Bourrier},
  {Demory}, {Fossati}, {Cubillos}, {Guenther}, {Kane}, {Kuschnig}, {Matthews},
  {Moffat}, {Rowe}, {Sasselov}, {Weiss}, \& {Winn}}]{sulis_2019}
{Sulis}, S., {Dragomir}, D., {Lendl}, M., {et~al.} 2019, \aap, 631, A129

\bibitem[{{Sur} {et~al.}(2020){Sur}, {Haskell}, \& {Kuhn}}]{sur_2020}
{Sur}, A., {Haskell}, B., \& {Kuhn}, E. 2020, \mnras, 495, 1360

\bibitem[{{Tremblin} \& {Chiang}(2013)}]{tremblin_2013}
{Tremblin}, P., \& {Chiang}, E. 2013, \mnras, 428, 2565

\bibitem[{{Tsiaras} {et~al.}(2016){Tsiaras}, {Rocchetto}, {Waldmann}, {Venot},
  {Varley}, {Morello}, {Damiano}, {Tinetti}, {Barton}, {Yurchenko}, \&
  {Tennyson}}]{tsiaras_2016}
{Tsiaras}, A., {Rocchetto}, M., {Waldmann}, I.~P., {et~al.} 2016, \apj, 820, 99

\bibitem[{{Valencia} {et~al.}(2010){Valencia}, {Ikoma}, {Guillot}, \&
  {Nettelmann}}]{valencia_2010}
{Valencia}, D., {Ikoma}, M., {Guillot}, T., \& {Nettelmann}, N. 2010, \aap,
  516, A20

\bibitem[{{van der Walt} {et~al.}(2011){van der Walt}, {Colbert}, \&
  {Varoquaux}}]{van_der_walt_2011}
{van der Walt}, S., {Colbert}, S.~C., \& {Varoquaux}, G. 2011, Computing in
  Science and Engineering, 13, 22

\bibitem[{{Virtanen} {et~al.}(2020){Virtanen}, {Gommers}, {Oliphant},
  {Haberland}, {Reddy}, {Cournapeau}, {Burovski}, {Peterson}, {Weckesser},
  {Bright}, {van der Walt}, {Brett}, {Wilson}, {Jarrod Millman}, {Mayorov},
  {Nelson}, {Jones}, {Kern}, {Larson}, {Carey}, {Polat}, {Feng}, {Moore}, {Vand
  erPlas}, {Laxalde}, {Perktold}, {Cimrman}, {Henriksen}, {Quintero}, {Harris},
  {Archibald}, {Ribeiro}, {Pedregosa}, {van Mulbregt}, \&
  {Contributors}}]{virtanen_2020}
{Virtanen}, P., {Gommers}, R., {Oliphant}, T.~E., {et~al.} 2020, Nature
  Methods, 17, 261

\bibitem[{{Vissapragada} {et~al.}(2020){Vissapragada}, {Knutson}, {Jovanovic},
  {Harada}, {Oklop{\v{c}}i{\'c}}, {Eriksen}, {Mawet}, {Millar-Blanchaer},
  {Tinyanont}, \& {Vasisht}}]{vissapragada_2020}
{Vissapragada}, S., {Knutson}, H.~A., {Jovanovic}, N., {et~al.} 2020, \aj, 159,
  278

\bibitem[{Wang \& Dai(2018)}]{wang_2018}
Wang, L., \& Dai, F. 2018, The Astrophysical Journal, 860, 175

\bibitem[{{Wang} \& {Dai}(2020)}]{wang_2020}
{Wang}, L., \& {Dai}, F. 2020, arXiv e-prints, arXiv:2101.00045

\bibitem[{{Zilinskas} {et~al.}(2020){Zilinskas}, {Miguel}, {Molli{\`e}re}, \&
  {Tsai}}]{zilinskas_2020}
{Zilinskas}, M., {Miguel}, Y., {Molli{\`e}re}, P., \& {Tsai}, S.-M. 2020,
  \mnras, 494, 1490

\end{thebibliography}

\end{document}